\documentclass{aastex631}

\usepackage{amsmath}

\begin{document}

\title{The First X-ray Polarization Observation of the Black Hole X-ray Binary 4U\,1630--47 \\ in the Steep Power Law State}

\author[0000-0001-5256-0278]{Nicole Rodriguez Cavero}
\affiliation{Physics Department, McDonnell Center for the Space Sciences, and Center for Quantum Leaps, Washington University in St. Louis, St. Louis, MO 63130, USA}

\author[0009-0001-4644-194X]{Lorenzo Marra}
\affiliation{Dipartimento di Matematica e Fisica, Universit\`a degli Studi Roma Tre, Via della Vasca Navale 84, 00146 Roma, Italy}

\author[0000-0002-1084-6507]{Henric Krawczynski}
\affiliation{Physics Department, McDonnell Center for the Space Sciences, and Center for Quantum Leaps, Washington University in St. Louis, St. Louis, MO 63130, USA}

\author[0000-0003-0079-1239]{Michal Dovčiak}
\affiliation{Astronomical Institute of the Czech Academy of Sciences, Boční II 1401/1, 14100 Praha 4, Czech Republic}

\author[0000-0002-4622-4240]{Stefano Bianchi}
\affiliation{Dipartimento di Matematica e Fisica, Universit\`a degli Studi Roma Tre, Via della Vasca Navale 84, 00146 Roma, Italy}

\author[0000-0002-5872-6061]{James F. Steiner}
\affiliation{Center for Astrophysics, Harvard \& Smithsonian, 60 Garden St, Cambridge, MA 02138, USA}

\author[0000-0003-2931-0742]{Jiri Svoboda}
\affiliation{Astronomical Institute of the Czech Academy of Sciences, Boční II 1401/1, 14100 Praha 4, Czech Republic}

\author[0000-0002-6384-3027]{Fiamma Capitanio}
\affiliation{INAF Istituto di Astrofisica e Planetologia Spaziali, Via del Fosso del Cavaliere 100, 00133 Roma, Italy}

\author[0000-0002-2152-0916]{Giorgio Matt}
\affiliation{Dipartimento di Matematica e Fisica, Universit\`a degli Studi Roma Tre, Via della Vasca Navale 84, 00146 Roma, Italy}

\author[0000-0002-6548-5622]{Michela Negro}
\affiliation{University of Maryland, Baltimore County, Baltimore, MD 21250, USA}
\affiliation{NASA Goddard Space Flight Center, Greenbelt, MD 20771, USA}
\affiliation{Center for Research and Exploration in Space Science and Technology, NASA/GSFC, Greenbelt, MD 20771, USA}

% ...




\begin{abstract}
The {\it Imaging X-ray Polarimetry Explorer} ({\it IXPE}) observed the black hole X-ray binary 4U\,1630--47
in the steep power law (or very high) state.
The observations reveal a linear polarization degree of the 2--8~keV X-rays
of $6.8 \pm 0.2\%$ at a position angle of 
$21\fdg3 \pm 0\fdg9$ East of North (all errors at
1$\sigma$ confidence level). 
Whereas the polarization degree increases with energy, the polarization angle stays constant within the accuracy of our measurements.
We compare the polarization of the source in the steep power-law state with the previous {\it IXPE} measurement of the source in the high soft state.
We find that even though the source flux and spectral shape are significantly different between the high soft state and the steep power-law state, their polarization signatures are similar.
Assuming that the polarization of 
both the thermal and power-law emission components are constant over time, we estimate the power-law component polarization to be 6.8--7.0\% and note that the polarization angle of the thermal and power-law components must be approximately aligned.
We discuss the implications for the origin of the power-law component and the properties of the emitting plasma.
\end{abstract}



\section{Introduction} \label{sec:intro}

Black hole X-ray binaries (BHXRBs) harbor a stellar mass black hole in close orbit with a companion star. The matter accreting onto the central black hole forms an accretion disk which is heated by internal frictions to the point of emitting radiation that typically peaks in the X-ray band.
BHXRB sources are found in different spectral states. The two main states, the high soft and low hard states (HSS and LHS, respectively), exhibit a spectrum that can be roughly described as a combination of both a soft thermal component and a harder electron-scattering component with reflection by a cold medium.
In the HSS, the X-rays are dominated by the thermal accretion disk emission followed by a non-thermal tail extending beyond $500$~keV. This state is often fitted with a multi-temperature blackbody model and a  power law $\propto E^{-\Gamma}$ with a photon index of $\Gamma \sim 2-2.2$ \citep{Zdziarski2004}.
In the LHS, the X-ray emission is dominated instead by
photons that Compton scatter in a hot coronal plasma, though a low-temperature disk component can still be detected \citep{2006csxs.book..157M}. In this state, BHXRB spectra consist of a cutoff power-law component with a typical photon index of $1.5 \leq \Gamma \leq 2.0$ and an exponential cutoff at high ($\sim 100$~keV) energies as well as reflected emission from the corona off the disk \citep{1991MNRAS.249..352G,2007A&ARv..15....1D}. 
BHXRBs can also be found in the steep power law (SPL) or very high state. The SPL state is characterized by competing thermal and power-law components---where the power-law component has a photon index of $\Gamma > 2.4$ (steeper than the higher energy tail of the HSS and the $\Gamma \sim 1.7$ detected in the LHS)
\citep{2006ARA&A..44...49R}.


The \textit{Imaging X-ray Polarimetry Explorer} \citep[\textit{IXPE},][]{ixpe} is a space-based observatory launched on 2021 December 9. 
\textit{IXPE} has measured the linear polarization of the 2--8~keV X-rays from several BHXRBs, giving new insights into the configuration and properties of their emitting plasmas. The \textit{IXPE} observations of the BHXRB Cyg X-1 in the LHS revealed a 4\% polarization aligned with the black hole radio jet, supporting the hypothesis that the jet might be launched from the black hole inner X-ray emitting region \citep{2022Sci...378..650K}. These results also revealed that the hot coronal plasma is extended parallel to the accretion disk plane and is seen at a higher inclination than the binary. 
\textit{IXPE} observed a high polarization degree of $\sim$20\% perpendicular to radio ejections of the black hole candidate Cyg X-3
suggesting that the primary source is inherently highly luminous but obscured so that only the reflected emission can be observed \citep{Veledina_2023}. 
The \textit{IXPE} observations of the low-inclination high-mass BHXRB LMC X-1 in the HSS gave only an upper limit on the total polarization degree of $<2.2\%$ \citep{2023arXiv230312034P} for a combination of two main spectral components: dominant thermal emission with a modest contribution of Comptonization. 

Observations of the transient low-mass X-ray binary (LMXRB) 4U\,1630--47 with the {\it Uhuru} satellite were first reported in \citet{1976ApJ...207L..25F} and \citet{1976ApJ...210L...9J}, describing four outbursts occurring every $\sim600$ days.
The X-ray spectral and timing properties of the LMXRB during an outburst in 1984 suggest the compact object of 4U\,1630--47 is a black hole candidate \citep{1986ApJ...304..664P} albeit with unusual outburst behavior \citep{2022MNRAS.510.1128C} indicative of a more complex system.
The source spectrum tends to show strong, blueshifted absorption lines corresponding to Fe XXV and Fe XXVI transitions during the soft accretion states \citep{2018ApJ...867...86P,2019MNRAS.482.2597G}. Previous measurements of the 4U\,1630--47 dust-scattering halo were used to estimate a distance range of 4.7--11.5~kpc \citep{2018ApJ...859...88K}. From the detection of short-duration dips in its X-ray light curve during outburst, a relatively high inclination of 60\degr--75\degr\ has been inferred \citep{1998ApJ...494..753K}.  Various reflection spectral modeling efforts have consistently measured a high spin: $a=0.985\substack{+0.005 \\ -0.014}$ \citep{2014ApJ...784L...2K}, $a=0.92 \pm 0.04$ \citep{2018ApJ...867...86P}, and $a\gtrsim0.9$ \citep{2021ApJ...909..146C}.

{\it IXPE} previously observed 4U\,1630--47 in the HSS where the detected emission was primarily from the thermal accretion disk \citep[][henceforth Paper\,I]{Ratheesh2023}. That observation revealed that the polarization degree increased with energy from $\sim$6\% at 2~keV to $\sim$10\% at 8~keV. 
The high  polarization degree and its energy dependence cannot be explained in terms of a standard geometrically thin accretion disk with a highly or fully  ionized accretion disk atmosphere \citep{Chandrasekhar1960,sob49,Sob63}. 
While a standard thin disk viewed at inclinations
$\gtrsim 85 \degr$ would produce a sufficiently high energy-integrated
polarization degree, relativistic effects would lead to a decrease of the polarization degree with energy contrary to the observed increase. Such a high inclination would also lead to eclipsing of the source which has not been detected.
In Paper\,I we argue that a geometrically thin disk with a partially ionized, outflowing emitting plasma can explain the observations.
The absorption in the emitting plasma leads to escaping emission that is likelier to have scattered only once and ends up being highly polarized parallel to the disk surface \citep{Losob79,LoskutovSobolev1981,2021MNRAS.501.3393T}. 
A vertically outflowing emitting plasma leads to increased emission angles in the local disk frame due to relativistic aberration resulting in a higher polarization degree \citep[e.g.][]{Beloborodov1998,Poutanen2023}.
Including absorption effects and the relativistic motion in the models achieves proper fits of the data for a thin accretion disk of a slowly spinning ($a \leq 0.5$) black hole seen at inclination $i \approx 75\degr$ when the emitting plasma has an optical thickness of $\tau \sim 7$ and moves with a vertical velocity $v \sim 0.5 \ c$.


\begin{figure} 
\epsscale{1.1}
\hspace*{-38mm}a)\hspace{0.475\textwidth}b)\\
\plottwo{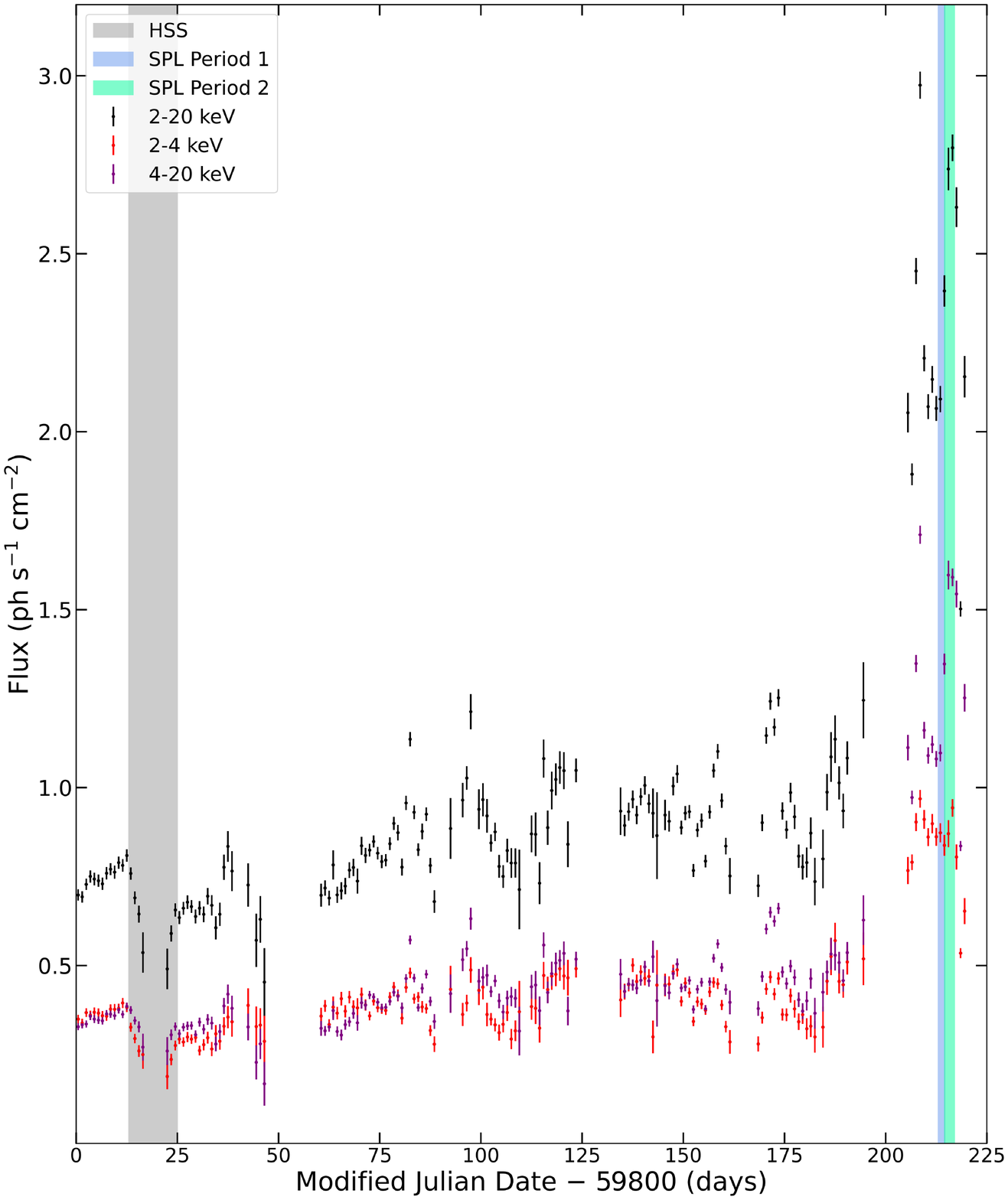}{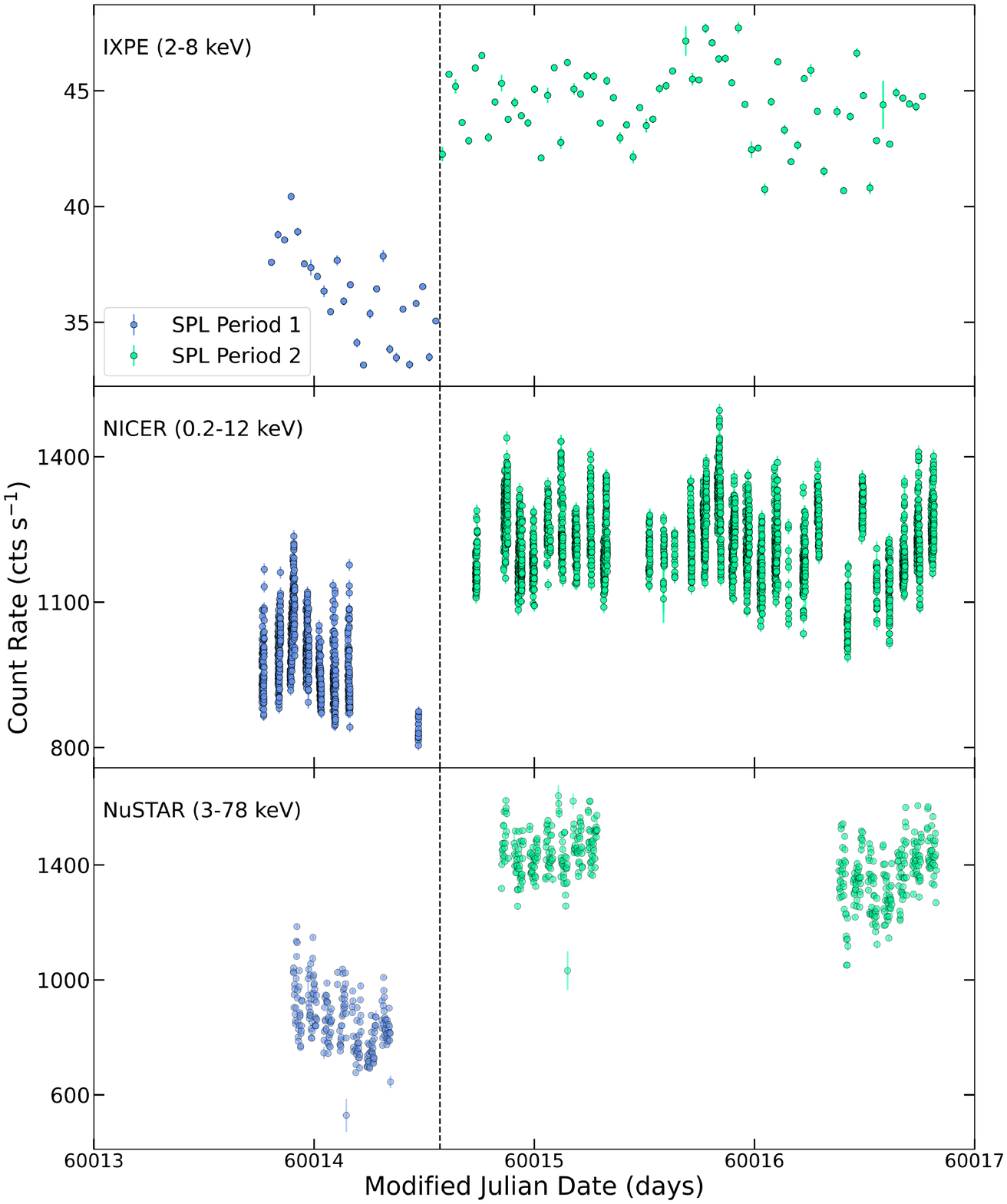}
\caption{X-ray light curves of 4U 1630--47. a) {MAXI} light curve between MJD 59800 (2022 August 9) and MJD 60025 (2023 March 22). The flux in the 2--20, 2--4 and 4--20 keV energy bands are reported in black, orange, and purple, respectively. The gray-shaded region corresponds to the observation reported in Paper\,I when the source was in the HSS while the regions shaded in blue (Period 1) and green (Period 2) correspond to the observation reported in this paper when the source was in the SPL state. b) From top to bottom: {\it IXPE}, NICER, and {\it NuSTAR} light curves from March 10 to March 14, 2023. Observations of Periods 1  and 2 are shown by the blue and green data points, respectively, with a sudden flux increase at around MJD 60014.57 indicated by the vertical dashed line.
\label{fig:lightcurves}}
\end{figure}

In this letter, we report on the first measurement of the polarization properties of a BHXRB in the SPL state. The letter is organized as follows. We describe the {\it IXPE}, NICER, and {\it NuSTAR}
observational results of 4U\,1630--47 in Section \ref{sec:results} and present a comparison of the polarization of the source in the HSS and the SPL states.
In Section \ref{sec:discussion}, we examine our results in the context of previous {\it IXPE} X-ray polarization measurements of BHXRBs and discuss scenarios that could explain the observed polarization signature.


\section{Data sets, Analysis Methods, and Results} \label{sec:results}

{\it IXPE} performed a target of opportunity (ToO) observation of 4U\,1630--47 between 2023 March 10 and 14 for $\sim$150~ks after daily monitoring of the source by the Gas Slit Camera (GSC) on the {Monitor of All-sky X-ray Image} ({MAXI}) \citep{maxi} reported a significant increase in flux, as shown in Figure \ref{fig:lightcurves}a.
The {MAXI} flux was about $0.62$~ph~s$^{-1}$~cm$^{-2}$ during the gray highlighted region of the figure which coincides with the Paper\,I observation---hereby referred to as the HSS data. The blue and green highlighted regions have a higher flux of approximately $2.24$~ph~s$^{-1}$~cm$^{-2}$ and $2.77$~ph~s$^{-1}$~cm$^{-2}$, respectively, signaling a change in the emission state of the source. During these later time intervals, the 4--20~keV flux shown in purple in Figure \ref{fig:lightcurves}a increases more drastically than the 2--4~keV flux shown in orange indicating an increase in the spectral hardness over this time.
Figure \ref{fig:lightcurves}b shows the {\it IXPE}, NICER \citep{nicer}, and {\it NuSTAR} \citep{nustar} 2--8, 0.2--12, and 3--78~keV count rates during these later intervals. We see the source flux increased dramatically around the time marked in the figure by the vertical dashed line: the {\it IXPE}, NICER, and {\it NuSTAR} count rates increased by $\sim 23\%$, $\sim 25\%$, and $\sim 63\%$, respectively.
The blanket {\it IXPE} coverage reveals that this increase was very sudden (about $\sim 2.6$~ks).
Owing to the drastic change, we divided our {\it IXPE}, NICER, and {\it NuSTAR} observations into Period 1 (blue) to Period 2 (green) before and after 13:42:53 UTC on 2023 March 11.
For a description of the {\it IXPE}, NICER, and {\it NuSTAR} data reduction, see Appendix~\ref{sec:datareduction}.

\begin{figure}
\epsscale{0.8}
\hspace*{4mm}a)\hspace{0.458\textwidth}b)\\
\includegraphics[width=0.43\textwidth]{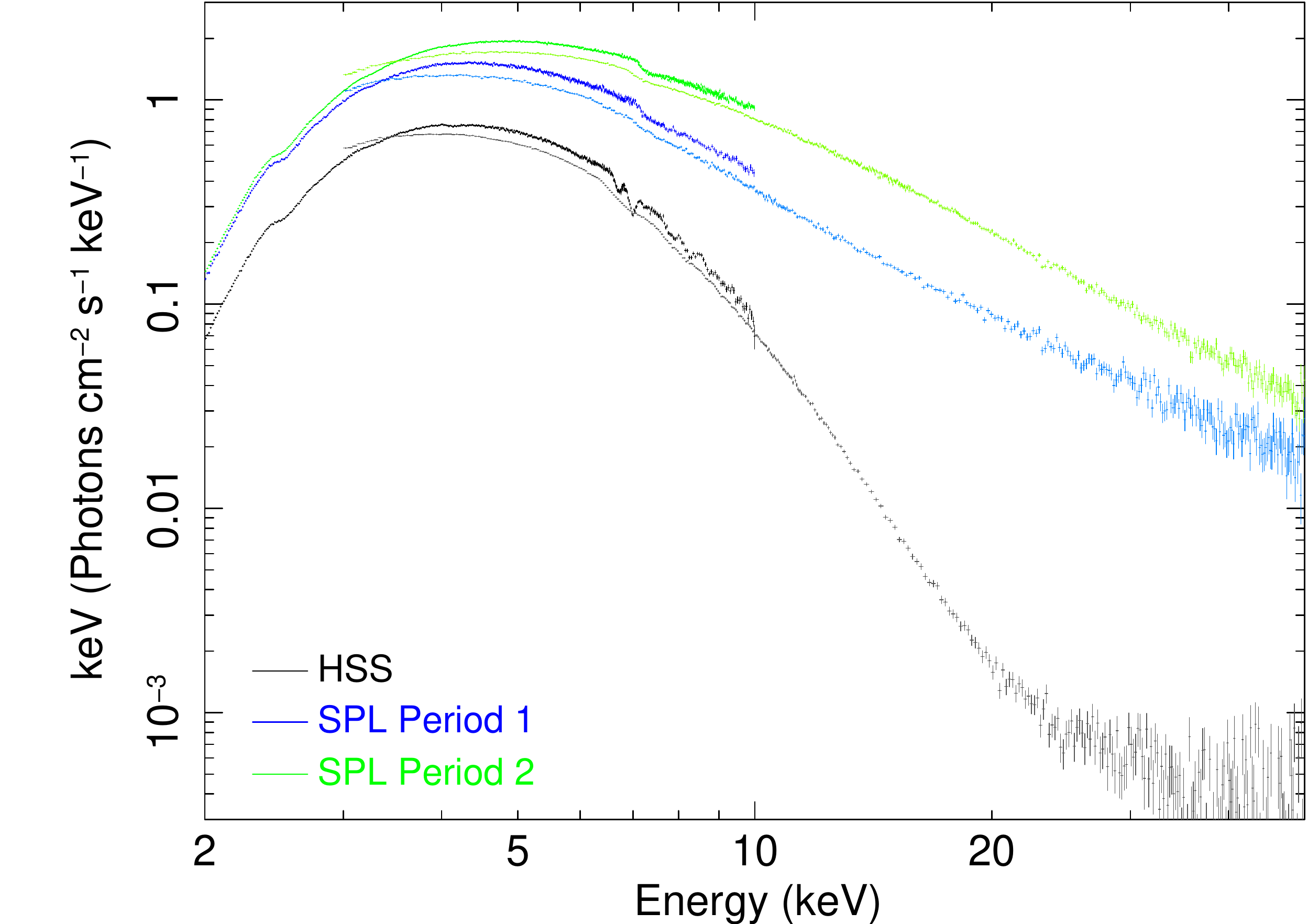}\hspace*{1cm}
\includegraphics[width=0.43\textwidth]{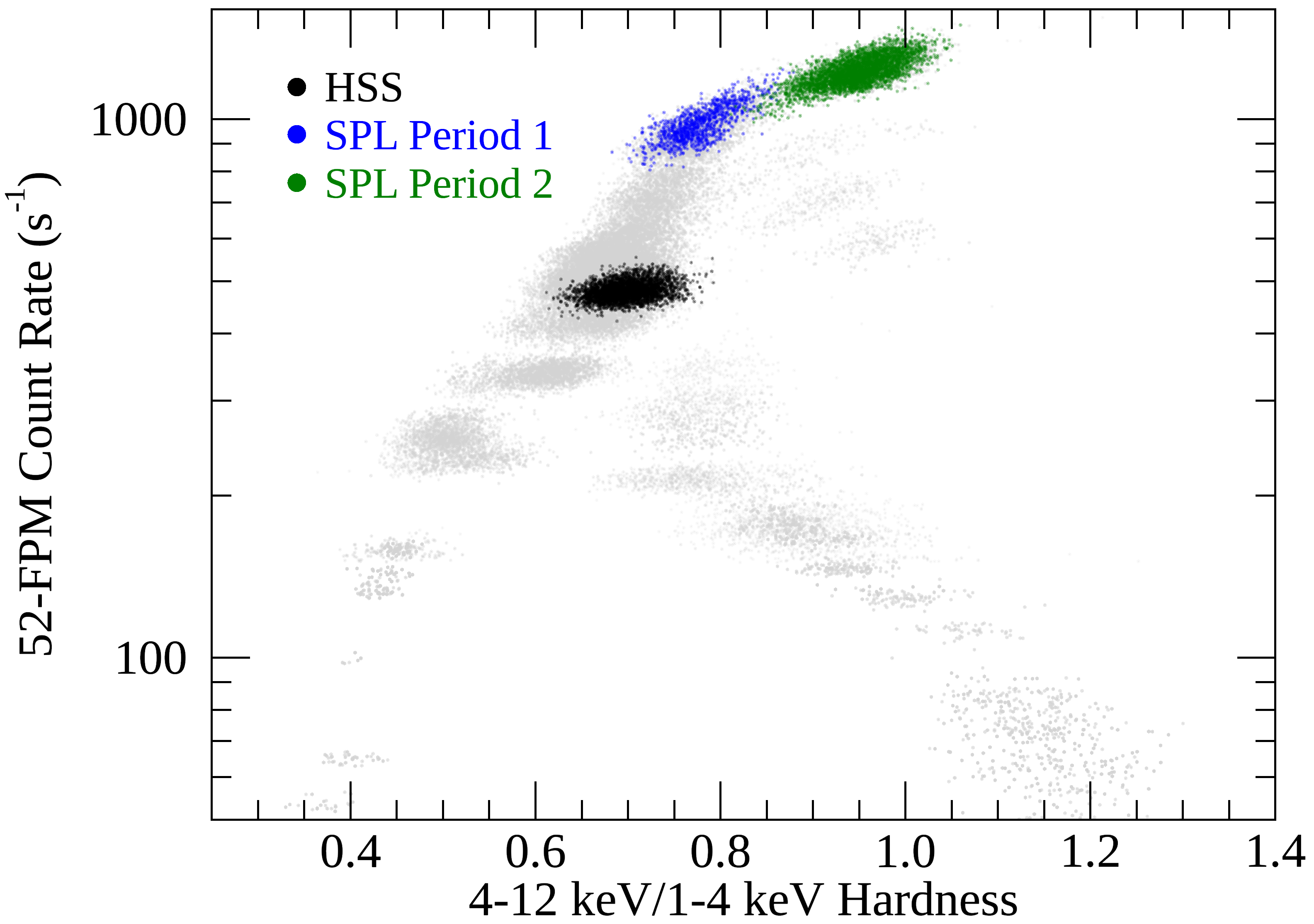}
\caption{a) NICER (2--10 keV) and {\it NuSTAR} (3--50 keV) spectra of the HSS (black) from Paper\,I and from the current SPL Period 1 (blue) and Period 2 observations (green). The spectra were unfolded using a unit constant model for both instruments. b) Hardness-intensity diagram from NICER data of the HSS (black) and SPL state Period 1 (blue) and Period 2 (green), in 8~s intervals.  Data from all previous NICER observations of 4U\,1630--47 are shown in gray. Rates have been normalized as if all 52 of NICER's FPMs were pointing at the source.
\label{fig:hardness}}
\end{figure}

A comparison of the NICER and {\it NuSTAR} spectra in Figure~\ref{fig:hardness}a for the HSS observation (black) and Periods 1 and~2 (blue and green) reveals that the source transitioned from the HSS to the SPL state. In Paper\,I, the power-law component of the spectra contributed $\sim 3 \%$ of the energy flux in the {\it IXPE} energy band. In contrast, our spectral fitting (see Appendix \ref{sec:spectralfit}) reveals that in Period 1 of the SPL state the power-law emission contributed $\sim$17--46\% of the 2--8\,keV emission while in Period 2 this contribution increased to $\sim$40--92\%. The soft HSS spectra are almost completely thermal in the form of a multi-temperature black body while the SPL spectra show an additional steep power-law component. From Figure~\ref{fig:hardness}a, we can see the SPL state shows an increase in 2--50~keV flux and a change in the spectral shape at energies above 5~keV. Only the HSS spectra exhibit prominent blueshifted Fe~XXV and Fe~XXVI lines as previously seen in past outbursts and explained in terms of over-ionization of the wind \citep{2014A&A...571A..76D} or of an intrinsic change of the physical properties of the wind itself \citep{2014ApJ...790...20H} in the SPL state.
Figure~\ref{fig:hardness}b shows a hardness--intensity diagram (HID) of 4U\,1630--47 NICER data including the HSS (black) and SPL (blue and green) observations contemporaneous with the {\it IXPE} measurements, and archival data. Period 2 exhibits the highest rate corresponding to the largest relative contribution of the power-law flux. The energy flux in the 1--12~keV band increases with hardness during the transition from the HSS to the SPL state saturating at $\sim 1496$~s$^{-1}$.
Most astrophysical black hole candidates move through a hardness-intensity diagram counter-clockwise during outbursts (see Figure 7 of \citealt{2004MNRAS.355.1105F} and Figure 1 of \citealt{2005Ap&SS.300..107H}). However, Figure \ref{fig:hardness}b shows  4U\,1630--47 evolving in a clockwise direction near the apex of the HID consistent with previous {\it Suzaku} observations of the source in the SPL state \citep{2014ApJ...790...20H}. We note that the variable motion of the source along the HID \citep[see Figure 11 of][]{2005ApJ...630..413T} makes it unclear if the source transitions from the HSS to the LHS through a high-intensity SPL regime or if we caught the source in an unusual pattern of motion. Furthermore, Figure \ref{fig:hardness}b shows no evident bright hard state, consistent with the results of \citet{2015MNRAS.450.3840C} which could indicate a deviation from the standard HID Q-track shape proposed in \citet{2004MNRAS.355.1105F}. Alternatively, \citet{2014ApJ...791...70T} suggest that a low large-scale magnetic field in the disk could delay the transition to the LHS.

\begin{figure} 
\epsscale{0.9}
\plotone{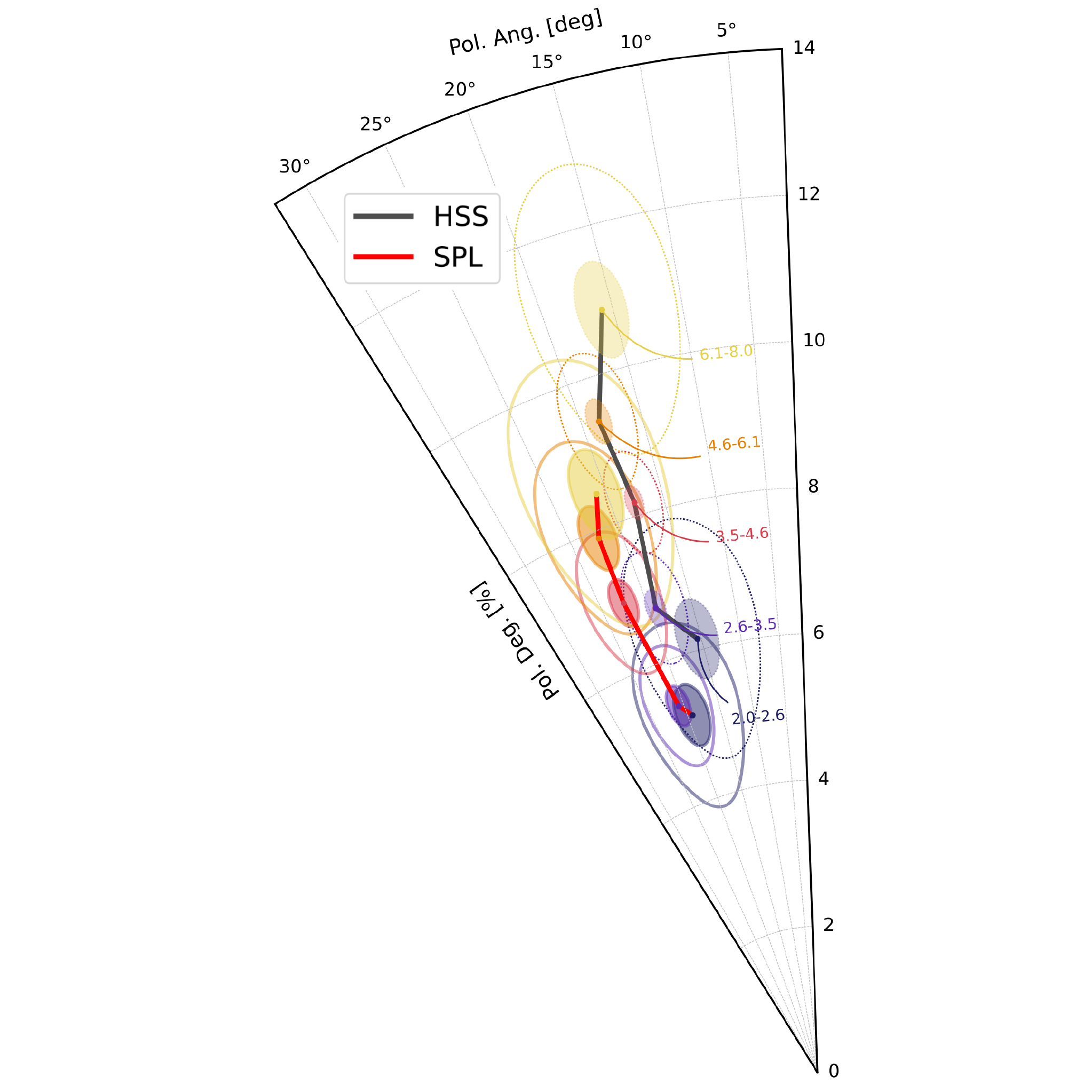}
\caption{Measured PD and PA of 4U\,1630--47 in 5 logarithmic energy bins: 2.0--2.6, 2.6--3.5, 3.5--4.6, 4.6--6.1, and 6.1--8.0~keV. The black line and transparent contours show the polarization in the HSS reported in Paper\,I. The red solid line and solid contours show the polarization in the SPL state (this paper). The shaded and unshaded ellipses show their $68.3\%$ and $99.7\%$ confidence regions, respectively.  Errors on PD and PA computed by {\sc ixpeobssim} are derived from the $Q$ and $U$ gaussian errors according to the formalism developed by \citet{2015APh....68...45K}.
\label{fig:polarplots}}
\end{figure}

\begin{figure}[ht!]
\epsscale{0.9}
\hspace*{-2.3cm}a)\hspace{0.44\textwidth}b)\\
\plottwo{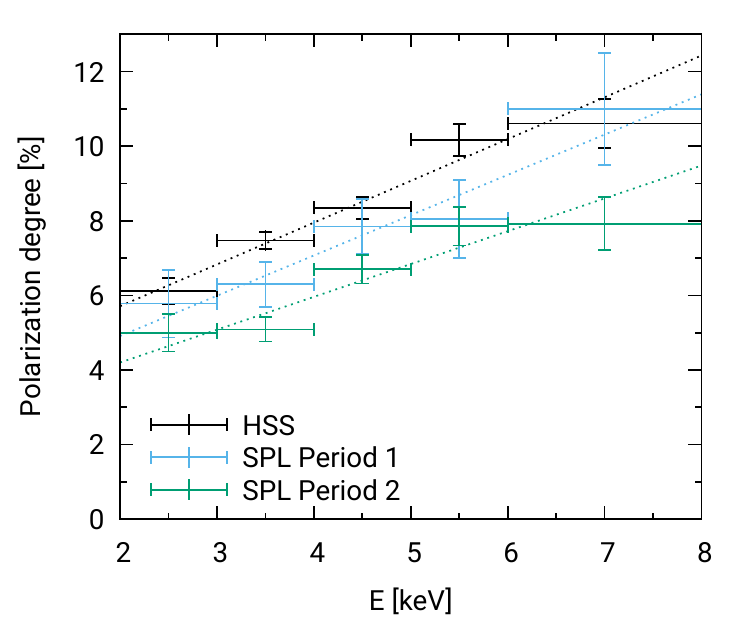}{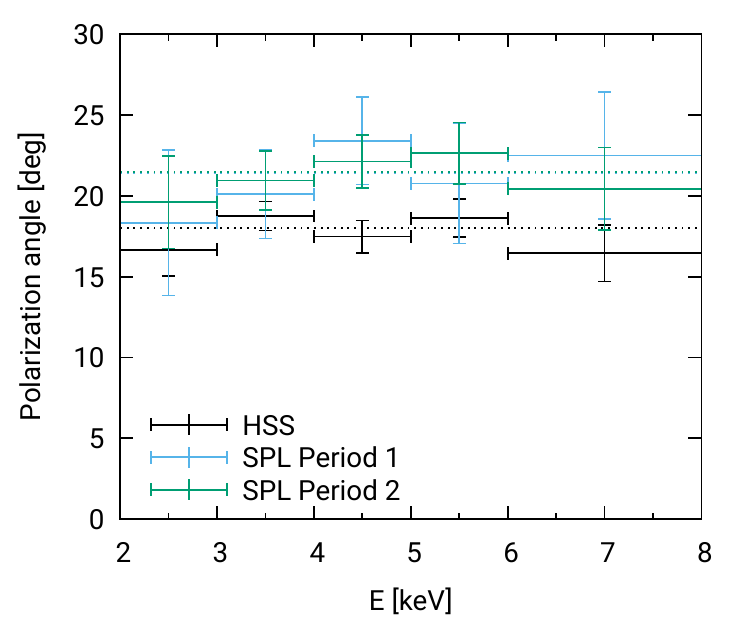}
\caption{
a) PD and b) PA as a function of energy in the {\it IXPE} 2--8~keV energy range. Comparison of the 4U\,1630--47 polarization properties in the HSS (black), reported in Paper\,I, and in the SPL Period 1 (blue) and Period 2 (green) discussed in this paper. Linear fits for PD and constant fit for PA are also shown in dotted lines (see the text for the fit details).
\label{fig:polvsenergy}}
\end{figure}

During the entire SPL state observation, {\it IXPE} measured an energy-averaged 2--8~keV linear polarization degree (PD) of $6.8 \pm 0.2 \%$ at a polarization angle (PA) $21\fdg3 \pm 0\fdg9$ (East of North) with a statistical confidence of over $30\,\sigma$. The SPL state observation has a $1.5 \%$ smaller PD than the $8.32\pm0.17\%$ HSS PD reported in Paper\,I at a PA $3\fdg5$ higher with respect to the previously observed $17\fdg8\pm 0\fdg6$.
Figure~\ref{fig:polarplots} shows the time-averaged polarization signature during both states in 5 logarithmic energy bands. 
The PA is constant within $3 \sigma$ during the HSS and SPL observations.
The summary of measured PD and PA in different spectral states is given in Table~\ref{tab:PD-PA}.
These values have been computed using the \verb|PCUBE| algorithm of the {\sc ixpeobssim} analysis software \citep{Baldini2022}. Figure~\ref{fig:polvsenergy} shows linear and constant fits of PD and PA, respectively, obtained using {\sc xspec} \citep{Arnaud1996}.
The HSS and SPL state observations have a similar linear dependence of the PD on the photon energy $E$, with a linear model ${\rm PD} = p_0 + \alpha(E/ 1\,\mathrm{keV})$. For the HSS, the reported values are $p_0 = 3.47 \pm 0.54 \%$, $\alpha = 1.12 \pm 0.13 \%$ with the null hypothesis probability of $3.55\times10^{-16}$ for a constant function.
For the SPL state Period 1 observation, these parameters change to $p_0 = 2.7 \pm 1.3 \%$, $\alpha = 1.08 \pm 0.32 \%$ with the null hypothesis probability of $1.42\times10^{-2}$ for a constant function.
For the SPL state Period 2 observation, these parameters are $p_0 = 2.44 \pm 0.70 \%$, $\alpha = 0.88 \pm 0.16 \%$ with the null hypothesis probability of $4.56\times10^{-7}$ for a constant function.
Both the HSS and SPL Period 1 and Period 2 observations show  relatively energy-independent PA in the {\it IXPE} band, with the fitted value of PA being $18\fdg0 \pm 0\fdg5$, $21\fdg4 \pm 1\fdg8$ and $21\fdg5 \pm 0\fdg9$ with the null hypothesis probability of 0.607, 0.854 and 0.877, respectively.

\begin{deluxetable}{ccccccc}
\tablewidth{0pt} 
\tablecaption{Polarization properties in different spectral states of 4U\,1630--47. The estimated fractions of the thermal and power-law flux contributing to the 2--8~keV energy band
are also given.}
\label{tab:PD-PA}
\tablehead{
\colhead{Spectral state} & \colhead{Polarization degree}& \colhead{Polarization angle} & \multicolumn{2}{c}{Thermal contribution} & \multicolumn{2}{c}{Power-law contribution}\\ 
& [\%] & [deg] &  Fit 1 &  Fit 2 & Fit 1 &  Fit 2}
\startdata 
HSS          & $8.32 \pm 0.17$ & $17.8 \pm 0.6$ & \multicolumn{2}{c}{0.97} & \multicolumn{2}{c}{0.03}\\
SPL Period 1 & $7.55 \pm 0.44$ & $21.7 \pm 1.7$ & 0.54 & 0.83 & 0.46 & 0.17\\
SPL Period 2 & $6.52 \pm 0.24$ & $21.3 \pm 1.1$ & 0.08 & 0.60 & 0.92 & 0.40\\
SPL Total    & $6.75 \pm 0.21$ & $21.3 \pm 0.9$ & \multicolumn{2}{c}{--} & \multicolumn{2}{c}{--}\\
\enddata
\tablecomments{Flux contributions are parameter-dependent. See Appendix \ref{sec:spectralfit} for more details on the model used. Contributions are calculated using either disk blackbody seed radiation (Fit 1) or blackbody seed radiation (Fit 2) for the power-law component of the spectra in the SPL Period 1 and 2 cases.}
\end{deluxetable}

\begin{figure} 
\epsscale{0.8}
\plotone{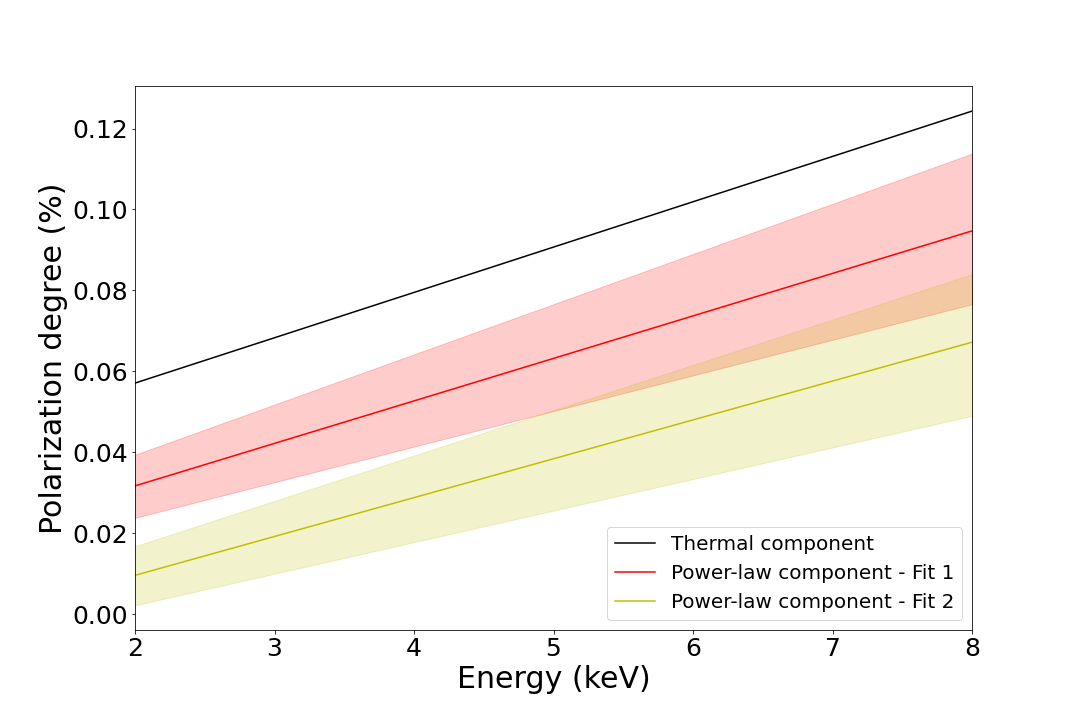}
\caption{Best linear fits with respect to energy of thermal component (black), power-law component for Fit 1 (red), and power-law component for Fit 2 (yellow). The shaded regions show the $1\sigma$ confidence intervals.
\label{fig:PDpowerlawcomp}}
\end{figure}

To study the polarization properties of the power-law component, we performed a polarimetric fit of the data starting from the spectral analysis described in Appendix \ref{sec:spectralfit}. We included the {\it IXPE} $Q$ and $U$ spectra in the spectral fit and convolved the thermal and power-law spectral components with two {\tt pollin} models\footnote{For a description of the linearly dependent polarization model see \url{https://heasarc.gsfc.nasa.gov/xanadu/xspec/manual/node213.html}. Note that these equations have been modified to the form described in the text.}. 
This allowed us to attribute polarization to each component separately assuming that the PD depends linearly on the photon energy $E$: PD~$=p_0+\alpha(E/1 \ \mathrm{keV})$. 
In Paper\,I, we found that the only spectral component contributing significantly to the HSS emission is the thermal one. We assumed that the polarization of this thermal component remains constant between the HSS and SPL states requiring that $p_{0_{\text{Thermal}}} = 3.47\%$ and $\alpha_{\text{Thermal}} = 1.12\%$ as per the HSS fit shown in Figure~\ref{fig:polvsenergy}a.
Due to the relatively constant PA during the HSS, SPL Period 1, and SPL Period 2 observations (Figure \ref{fig:polvsenergy}b), we further assumed that the thermal and non-thermal components have equal PA and allowed it to vary between SPL periods. Additionally, the PA appears to be energy-independent so our fits take the PA to be constant with energy: PA$=\psi$.
As shown in Table \ref{tab:PD-PA}, the estimates of the power-law component flux contribution depend on the model parameters used and will therefore also affect the estimate of the polarization properties of the power-law component.
Figure \ref{fig:PDpowerlawcomp} summarizes the results of our linear fits for the non-thermal component PD resulting from Fits 1 and 2 as well as the assumed thermal component PD for comparison.
For Fit 1, we assumed a multi-color blackbody as the Comptonized component input radiation (Figure \ref{fig:specfits}a). For the PD of the power-law component, we found that ${\alpha_{\text{Fit1}}} = 1.05 \pm  0.45 \%$ and we set an upper limit on ${p_{0_{\text{Fit1}}}}$ of $2.7 \%$.
The ${p_{0_{\text{Fit1}}}}$ upper limit tells us that the Comptonization component could be unpolarized at 0~keV but this is just an extrapolation---the power-law PD in the 2--8~keV energy range (Figure \ref{fig:PDpowerlawcomp}) shows that the component is polarized.
The computed PAs for Period 1 and Period 2 are ${\psi_{\text{Fit1--P1}}} = 21\fdg0 \pm 3\fdg4$ and ${\psi_{\text{Fit1--P2}}} = 21\fdg7 \pm 2\fdg2$.
For Fit 2 (Figure \ref{fig:specfits}b), we assume a simple blackbody as a seed for the power-law radiation. In this case, the thermal emission is the main source of flux in the 2--8~keV energy range for both Periods 1 and 2. The PD of the power-law component  can be fitted with ${\alpha_{\text{Fit2}}} = 0.96 \pm  0.26 \%$ and we were only able to set an upper limit on ${p_{0_{\text{Fit2}}}}$ of $1.3 \%$. The corresponding PAs for Period 1 and Period 2 are ${\psi_{\text{Fit2--P1}}} = 21\fdg0 \pm 3\fdg5$ and ${\psi_{\text{Fit2--P2}}} = 21\fdg7 \pm 2\fdg1$.
We also calculated the 2--8~keV average PD of the power-law component from the {\it IXPE} $I$, $Q$, and $U$ fluxes. For Fit~1, we get $7.0 \pm 3.2 \%$ and $6.8 \pm 2.6 \%$ in Periods 1 and 2, respectively. For Fit~2, we get $6.8 \pm 3.9\%$ and $7.0 \pm 2.2 \%$ in Periods 1 and 2, respectively.

\section{Discussion}
\label{sec:discussion}

\textit{IXPE} observed 4U 1630--47 in the HSS (Paper I) and in the SPL state (this paper).
We find that the HSS and SPL exhibit surprisingly similar polarization properties despite their very different energy spectra. Although the PD of the HSS (increasing from 6\% to 10\% between 2 to 8 keV) exceeded that of the SPL observations (increasing from 5\%  to 8\% between 2 to 8 keV), and  Figure~\ref{fig:polvsenergy}a shows that the PD of Period~2 decreases with respect to Period~1, 
we note that the PD varied as much during the HSS observations (Fig.~M3 of Paper\,I) as it did between the HSS and the SPL observations. 
The change in polarization direction $\sim3\fdg5$ is not statistically significant ($3 \sigma$).
While the HSS spectrum was dominated by the thermal component, our spectral analysis shows that the Comptonization component increased by a large factor between the HSS, SPL Period 1 and Period 2, although its exact flux contribution is model parameter-dependent.
Since the polarization angle stays almost the same with vastly different flux contributions of the power-law component, this component has to be polarized in a similar direction as the thermal component. Our polarimetric analysis reveals that the power-law component has an energy-integrated PD of 6.8--7.0\% in both cases analyzed, i.e. using either multicolor disk blackbody or single temperature blackbody as seed photons for Comptonization. Since both cases suggest substantially different contributions of this component to the total flux, we consider this estimate to be quite independent of the model assumptions. Note that the dominating thermal component in HSS had a PD of 8.3 \%, thus the Comptonized component is slightly less polarized than the thermal one by approximately 1.3--1.5\%.

This congruence of the PD and directions is puzzling if the emission comes from spatially distinct regions and is produced by different physical emission mechanisms.
Direct thermal emission from the disk tends to be polarized parallel to the accretion disk except for close to the innermost stable circular orbit (ISCO) where strong gravitational effects rotate the PA by about 10\degr\  \citep{1977Natur.269..128C,Loktev22}. Gravitationally lensed photons that scatter off the disk (known as returning radiation) are polarized perpendicular to the direct thermal radiation \citep{2009ApJ...701.1175S}.
Comptonization, commonly invoked to explain the power-law component, gives rise to a polarization perpendicular to the spatial extent of the Comptonizing plasma \citep{PS96,2010ApJ...712..908S,2022ApJ...934....4K}. 
The apparent alignment of the polarization directions of the thermal and power-law  emission could imply that the Comptonizing plasma of the SPL state is extended perpendicular to the accretion disk---contrary to what we inferred for the hard state of Cyg X-1 \citep{2022Sci...378..650K}.
However, it is worth noting that for a slab corona geometry, polarization is parallel to the disk at photon energies where the first Compton scattering dominates the flux \citep{Poutanen2023}. Since the temperature of the disk is high ($kT_\textrm{bb} \approx 1.5$~keV), the first scattering could dominate in the {\it IXPE} energy range such that the PA of the disk and the up-scattered component are aligned.

Based on the \textit{IXPE} results, we posit that the HSS and SPL states could exhibit similar disk geometries and involve similar emission processes.
In the scenario discussed in Paper\,I, an outflowing, partially-ionized accretion disk atmosphere produces the observed high PD as a result of Thomson scattering.
The electrons in the outflow attain Compton temperature (a few keV) if efficient heating and acceleration mechanisms, such as shocks, magnetic reconnection, and turbulence, do not operate.
Instantaneous increase of electron heating and acceleration may lead to a change of the scattering mechanism---from Thomson to inverse Compton---producing the observed power-law component.
During the transitions between the soft and hard states, the observed spectra are known to be well fitted with Comptonization from low-temperature thermal or hybrid (thermal and non-thermal) electrons \citep{Gierlinski1999,Zdziarski2001,Zycki2001}, with a typical temperature of the Maxwellian part $\sim10$~keV.
Increased electron temperature, in general, causes the reduction of the PD \citep[e.g., Fig. 2 of][]{Poutanen1994}; however for these low electron temperatures the effect is rather small and the polarization signatures remain similar to (albeit not exactly the same as in) the Thomson-scattering case.
The observed variations of the PD during the HSS and SPL states could result from changes in the scattered fraction and/or the outflow velocities.

As mentioned in Paper\,I and in \citet{West23}, non-vanishing accretion disk geometrical thicknesses may play a role in explaining the high polarization fractions of the source.
Spectral fitting indicates that the disk temperature $kT_\textrm{bb}$ increased between the HSS (in Paper\,I) and the SPL state. This increase in temperature is expected if a thicker accretion disk is present in the SPL state \citep{2005ApJ...630..413T}.
As higher energy photons originate closer to the black hole and are more likely to scatter, this scenario naturally explains PD increasing with photon energy. In contrast, the reflection off distant features (e.g. off a wind) would give rise to rather energy-independent PD. We also note that the neutral hydrogen column density is much smaller in the SPL than in the HSS state. The similar polarization properties of the emission from both states confirm our conclusion from Paper I that scattering off the wind is most likely not the dominant mechanism explaining the high polarization of the X-ray emission. 

On the other hand, we note that spectral timing studies of black hole LMXRBs suggest that their coronae contract in the hard state and then expand during the hard-to-soft state transition \citep{2022ApJ...930...18W}. Soft reverberation lag modeling employing a lamppost corona estimates that the corona height increases by an order of magnitude during the state transition \citep{2021ApJ...910L...3W}. If this increase in height were to be accompanied by a decrease in width, we could expect a change in the shape of the corona from laterally extended in the LHS to vertically extended---and hence giving rise to large reverberation lags---in the intermediate states. Our polarization results could then be explained by a cone or lamppost-shaped corona in the SPL state. Future polarization measurements of the source, particularly in the LHS, could help constrain the evolution of the corona geometry as well as the polarization of the power-law component.

In other alternative scenarios, the power-law component could originate as synchrotron emission from a jet perpendicular to the accretion disk threaded by a magnetic field aligned with the jet; or from synchrotron emission from non-thermal electrons accelerated in the plunging region, gyrating in a magnetic field perpendicular to the accretion disk \citep{2022MNRAS.515..775H}. This model would require just the right amount of magnetic field non-uniformity to explain the rather low PD of the power-law emission for synchrotron emission. \citet{2011RAA....11..631Y} propose that the SPL state originates from synchrotron radiation of magnetized compact spots near the ISCO, down-scattered by thermal electrons in the corona. Also here, some fine-tuning is required so that the combined thermal and power-law emission end up having similar polarization signatures as the thermal emission alone.

\begin{acknowledgments}
{\it Acknowledgments:}
The {\it IXPE} is a joint US and Italian mission. The US contribution is supported by NASA and led and managed by its Marshall Space Flight Center (MSFC), with industry partner Ball Aerospace (contract NNM15AA18C). The Italian contribution is supported by the Italian Space agency (Agenzia Spaziale Italiana, ASI) through contract ASI-OHBI-2017-12-I.0, agreements ASI-INAF-2017-12-H0 and ASI-INFN-2017.13-H0, and its Space Science Data Center (SSDC), and by the Istituto Nazionale di Astrofisica (INAF) and the Istituto Nazionale di Fisica Nucleare (INFN) in Italy. This research used data and software products or online services provided by the {\it IXPE} Team (Marshall Space Flight Center, the Space Science Data Center of the Italian Space Agency, the Istituto Nazionale di Astrofisica, and Istituto Nazionale di Fisica Nucleare), as well as the High-Energy Astrophysics Science Archive Research Center (HEASARC), at NASA Goddard Space Flight Center.
N.R.C and H.K acknowledge NASA support through the grants NNX16AC42G, 80NSSC20K0329, 80NSSC20K0540, NAS8- 03060, 80NSSC21K1817, 80NSSC22K1291, and 80NSSC22K1883 as well as support from the McDonnell Center for the Space Sciences.
M.D., J.Sv., V.K. and J.Pod. acknowledge the support from the Czech Science Foundation project GACR 21-06825X and the institutional support  from the Astronomical Institute of the Czech Academy of Sciences RVO:67985815.
A.I. acknowledges support from the Royal Society. 
M.N. acknowledges the support by NASA under award number 80GSFC21M0002. POP acknowledges financial support from the French High Energy Programme (PNHE) of the CNRS as well as from the French Space Agency (CNES).
\end{acknowledgments}

\bibliography{4U163-47-ApJLetter}{}

\begin{thebibliography}{}
\expandafter\ifx\csname natexlab\endcsname\relax\def\natexlab#1{#1}\fi
\providecommand{\url}[1]{\href{#1}{#1}}
\providecommand{\dodoi}[1]{doi:~\href{http://doi.org/#1}{\nolinkurl{#1}}}
\providecommand{\doeprint}[1]{\href{http://ascl.net/#1}{\nolinkurl{http://ascl.net/#1}}}
\providecommand{\doarXiv}[1]{\href{https://arxiv.org/abs/#1}{\nolinkurl{https://arxiv.org/abs/#1}}}

\bibitem[{{Arnaud}(1996)}]{Arnaud1996}
{Arnaud}, K.~A. 1996, in ASP Conf. Ser., Vol. 101, Astronomical Data Analysis
  Software and Systems V, ed. G.~H. {Jacoby} \& J.~{Barnes} (San Francisco:
  Astron. Soc. Pac.), 17--20

\bibitem[{{Baldini} {et~al.}(2022){Baldini}, {Bucciantini}, {Lalla}, {Ehlert},
  {Manfreda}, {Negro}, {Omodei}, {Pesce-Rollins}, {Sgr{\`o}}, \&
  {Silvestri}}]{Baldini2022}
{Baldini}, L., {Bucciantini}, N., {Lalla}, N.~D., {et~al.} 2022, SoftwareX, 19,
  101194, \dodoi{10.1016/j.softx.2022.101194}

\bibitem[{{Beloborodov}(1998)}]{Beloborodov1998}
{Beloborodov}, A.~M. 1998, \apjl, 496, L105, \dodoi{10.1086/311260}

\bibitem[{{Capitanio} {et~al.}(2015){Capitanio}, {Campana}, {De Cesare}, \&
  {Ferrigno}}]{2015MNRAS.450.3840C}
{Capitanio}, F., {Campana}, R., {De Cesare}, G., \& {Ferrigno}, C. 2015,
  \mnras, 450, 3840, \dodoi{10.1093/mnras/stv687}

\bibitem[{Chandrasekhar(1960)}]{Chandrasekhar1960}
Chandrasekhar, S. 1960, {R}adiative {T}ransfer (New York: Dover Publications)

\bibitem[{{Chatterjee} {et~al.}(2022){Chatterjee}, {Debnath}, {Bhowmick},
  {Nath}, \& {Chatterjee}}]{2022MNRAS.510.1128C}
{Chatterjee}, K., {Debnath}, D., {Bhowmick}, R., {Nath}, S.~K., \&
  {Chatterjee}, D. 2022, \mnras, 510, 1128, \dodoi{10.1093/mnras/stab3570}

\bibitem[{{Connors} \& {Stark}(1977)}]{1977Natur.269..128C}
{Connors}, P.~A., \& {Stark}, R.~F. 1977, \nat, 269, 128,
  \dodoi{10.1038/269128a0}

\bibitem[{{Connors} {et~al.}(2021){Connors}, {Garc{\'\i}a}, {Tomsick}, {Hare},
  {Dauser}, {Grinberg}, {Steiner}, {Mastroserio}, {Sridhar}, {Fabian}, {Jiang},
  {Parker}, {Harrison}, \& {Kallman}}]{2021ApJ...909..146C}
{Connors}, R. M.~T., {Garc{\'\i}a}, J.~A., {Tomsick}, J., {et~al.} 2021, \apj,
  909, 146, \dodoi{10.3847/1538-4357/abdd2c}

\bibitem[{{Di Marco} {et~al.}(2023){Di Marco}, {Soffitta}, {Costa},
  {Ferrazzoli}, {La Monaca}, {Rankin}, {Ratheesh}, {Xie}, {Baldini}, {Del
  Monte}, {Ehlert}, {Fabiani}, {Kim}, {Muleri}, {O'Dell}, {Ramsey}, {Rubini},
  {Sgr{\`o}}, {Silvestri}, {Tennant}, \& {Weisskopf}}]{2023AJ....165..143D}
{Di Marco}, A., {Soffitta}, P., {Costa}, E., {et~al.} 2023, \aj, 165, 143,
  \dodoi{10.3847/1538-3881/acba0f}

\bibitem[{{D{\'\i}az Trigo} {et~al.}(2014){D{\'\i}az Trigo}, {Migliari},
  {Miller-Jones}, \& {Guainazzi}}]{2014A&A...571A..76D}
{D{\'\i}az Trigo}, M., {Migliari}, S., {Miller-Jones}, J.~C.~A., \&
  {Guainazzi}, M. 2014, \aap, 571, A76, \dodoi{10.1051/0004-6361/201424554}

\bibitem[{{Done} {et~al.}(2007){Done}, {Gierli{\'n}ski}, \&
  {Kubota}}]{2007A&ARv..15....1D}
{Done}, C., {Gierli{\'n}ski}, M., \& {Kubota}, A. 2007, \aapr, 15, 1,
  \dodoi{10.1007/s00159-007-0006-1}

\bibitem[{{Fender} {et~al.}(2004){Fender}, {Belloni}, \&
  {Gallo}}]{2004MNRAS.355.1105F}
{Fender}, R.~P., {Belloni}, T.~M., \& {Gallo}, E. 2004, \mnras, 355, 1105,
  \dodoi{10.1111/j.1365-2966.2004.08384.x}

\bibitem[{{Ferland} {et~al.}(2017){Ferland}, {Chatzikos}, {Guzm{\'a}n},
  {Lykins}, {van Hoof}, {Williams}, {Abel}, {Badnell}, {Keenan}, {Porter}, \&
  {Stancil}}]{2017RMxAA..53..385F}
{Ferland}, G.~J., {Chatzikos}, M., {Guzm{\'a}n}, F., {et~al.} 2017, \rmxaa, 53,
  385, \dodoi{10.48550/arXiv.1705.10877}

\bibitem[{{Forman} {et~al.}(1976){Forman}, {Jones}, \&
  {Tananbaum}}]{1976ApJ...207L..25F}
{Forman}, W., {Jones}, C., \& {Tananbaum}, H. 1976, \apjl, 207, L25,
  \dodoi{10.1086/182170}

\bibitem[{{Gatuzz} {et~al.}(2019){Gatuzz}, {D{\'\i}az Trigo}, {Miller-Jones},
  \& {Migliari}}]{2019MNRAS.482.2597G}
{Gatuzz}, E., {D{\'\i}az Trigo}, M., {Miller-Jones}, J.~C.~A., \& {Migliari},
  S. 2019, \mnras, 482, 2597, \dodoi{10.1093/mnras/sty2850}

\bibitem[{{Gendreau} {et~al.}(2012){Gendreau}, {Arzoumanian}, \&
  {Okajima}}]{nicer}
{Gendreau}, K.~C., {Arzoumanian}, Z., \& {Okajima}, T. 2012, in \procspie, Vol.
  8443, Space Telescopes and Instrumentation 2012: Ultraviolet to Gamma Ray,
  ed. T.~{Takahashi}, S.~S. {Murray}, \& J.-W.~A. {den Herder}, 844313,
  \dodoi{10.1117/12.926396}

\bibitem[{{George} \& {Fabian}(1991)}]{1991MNRAS.249..352G}
{George}, I.~M., \& {Fabian}, A.~C. 1991, \mnras, 249, 352,
  \dodoi{10.1093/mnras/249.2.352}

\bibitem[{{Gierli{\'n}ski} {et~al.}(1999){Gierli{\'n}ski}, {Zdziarski},
  {Poutanen}, {Coppi}, {Ebisawa}, \& {Johnson}}]{Gierlinski1999}
{Gierli{\'n}ski}, M., {Zdziarski}, A.~A., {Poutanen}, J., {et~al.} 1999,
  \mnras, 309, 496, \dodoi{10.1046/j.1365-8711.1999.02875.x}

\bibitem[{{Hankla} {et~al.}(2022){Hankla}, {Scepi}, \&
  {Dexter}}]{2022MNRAS.515..775H}
{Hankla}, A.~M., {Scepi}, N., \& {Dexter}, J. 2022, \mnras, 515, 775,
  \dodoi{10.1093/mnras/stac1785}

\bibitem[{{Harrison} {et~al.}(2013){Harrison}, {Craig}, {Christensen},
  {Hailey}, {Zhang}, {Boggs}, {Stern}, {Cook}, {Forster}, {Giommi},
  {Grefenstette}, {Kim}, {Kitaguchi}, {Koglin}, {Madsen}, {Mao}, {Miyasaka},
  {Mori}, {Perri}, {Pivovaroff}, {Puccetti}, {Rana}, {Westergaard}, {Willis},
  {Zoglauer}, {An}, {Bachetti}, {Barri{\`e}re}, {Bellm}, {Bhalerao},
  {Brejnholt}, {Fuerst}, {Liebe}, {Markwardt}, {Nynka}, {Vogel}, {Walton},
  {Wik}, {Alexander}, {Cominsky}, {Hornschemeier}, {Hornstrup}, {Kaspi},
  {Madejski}, {Matt}, {Molendi}, {Smith}, {Tomsick}, {Ajello}, {Ballantyne},
  {Balokovi{\'c}}, {Barret}, {Bauer}, {Blandford}, {Niel Brandt}, {Brenneman},
  {Chiang}, {Chakrabarty}, {Chenevez}, {Comastri}, {Dufour}, {Elvis}, {Fabian},
  {Farrah}, {Fryer}, {Gotthelf}, {Grindlay}, {Helfand}, {Krivonos}, {Meier},
  {Miller}, {Natalucci}, {Ogle}, {Ofek}, {Ptak}, {Reynolds}, {Rigby},
  {Tagliaferri}, {Thorsett}, {Treister}, \& {Urry}}]{nustar}
{Harrison}, F.~A., {Craig}, W.~W., {Christensen}, F.~E., {et~al.} 2013, \apj,
  770, 103, \dodoi{10.1088/0004-637X/770/2/103}

\bibitem[{{Homan} \& {Belloni}(2005)}]{2005Ap&SS.300..107H}
{Homan}, J., \& {Belloni}, T. 2005, \apss, 300, 107,
  \dodoi{10.1007/s10509-005-1197-4}

\bibitem[{{Hori} {et~al.}(2014){Hori}, {Ueda}, {Shidatsu}, {Kawamuro},
  {Kubota}, {Done}, {Nakahira}, {Tsumura}, {Shirahata}, \&
  {Nagayama}}]{2014ApJ...790...20H}
{Hori}, T., {Ueda}, Y., {Shidatsu}, M., {et~al.} 2014, \apj, 790, 20,
  \dodoi{10.1088/0004-637X/790/1/20}

\bibitem[{{Jones} {et~al.}(1976){Jones}, {Forman}, {Tananbaum}, \&
  {Turner}}]{1976ApJ...210L...9J}
{Jones}, C., {Forman}, W., {Tananbaum}, H., \& {Turner}, M.~J.~L. 1976, \apjl,
  210, L9, \dodoi{10.1086/182291}

\bibitem[{{Joye} \& {Mandel}(2003)}]{ds9}
{Joye}, W.~A., \& {Mandel}, E. 2003, in ASP Conf. Ser., Vol. 295, Astronomical
  Data Analysis Software and Systems XII, ed. H.~E. {Payne}, R.~I.
  {Jedrzejewski}, \& R.~N. {Hook} (San Francisco: Astron. Soc. Pac.), 489

\bibitem[{{Kalemci} {et~al.}(2018){Kalemci}, {Maccarone}, \&
  {Tomsick}}]{2018ApJ...859...88K}
{Kalemci}, E., {Maccarone}, T.~J., \& {Tomsick}, J.~A. 2018, \apj, 859, 88,
  \dodoi{10.3847/1538-4357/aabcd3}

\bibitem[{{King} {et~al.}(2014){King}, {Walton}, {Miller}, {Barret}, {Boggs},
  {Christensen}, {Craig}, {Fabian}, {F{\"u}rst}, {Hailey}, {Harrison},
  {Krivonos}, {Mori}, {Natalucci}, {Stern}, {Tomsick}, \&
  {Zhang}}]{2014ApJ...784L...2K}
{King}, A.~L., {Walton}, D.~J., {Miller}, J.~M., {et~al.} 2014, \apjl, 784, L2,
  \dodoi{10.1088/2041-8205/784/1/L2}

\bibitem[{{Kislat} {et~al.}(2015){Kislat}, {Clark}, {Beilicke}, \&
  {Krawczynski}}]{2015APh....68...45K}
{Kislat}, F., {Clark}, B., {Beilicke}, M., \& {Krawczynski}, H. 2015,
  Astroparticle Physics, 68, 45, \dodoi{10.1016/j.astropartphys.2015.02.007}

\bibitem[{{Krawczynski} \& {Beheshtipour}(2022)}]{2022ApJ...934....4K}
{Krawczynski}, H., \& {Beheshtipour}, B. 2022, \apj, 934, 4,
  \dodoi{10.3847/1538-4357/ac7725}

\bibitem[{{Krawczynski} {et~al.}(2022){Krawczynski}, {Muleri}, {Dov{\v{c}}iak},
  {Veledina}, {Rodriguez Cavero}, {Svoboda}, {Ingram}, {Matt}, {Garcia},
  {Loktev}, {Negro}, {Poutanen}, {Kitaguchi}, {Podgorn{\'y}}, {Rankin},
  {Zhang}, {Berdyugin}, {Berdyugina}, {Bianchi}, {Blinov}, {Capitanio}, {Di
  Lalla}, {Draghis}, {Fabiani}, {Kagitani}, {Kravtsov}, {Kiehlmann},
  {Latronico}, {Lutovinov}, {Mandarakas}, {Marin}, {Marinucci}, {Miller},
  {Mizuno}, {Molkov}, {Omodei}, {Petrucci}, {Ratheesh}, {Sakanoi}, {Semena},
  {Skalidis}, {Soffitta}, {Tennant}, {Thalhammer}, {Tombesi}, {Weisskopf},
  {Wilms}, {Zhang}, {Agudo}, {Antonelli}, {Bachetti}, {Baldini}, {Baumgartner},
  {Bellazzini}, {Bongiorno}, {Bonino}, {Brez}, {Bucciantini}, {Castellano},
  {Cavazzuti}, {Ciprini}, {Costa}, {De Rosa}, {Del Monte}, {Di Gesu}, {Di
  Marco}, {Donnarumma}, {Doroshenko}, {Ehlert}, {Enoto}, {Evangelista},
  {Ferrazzoli}, {Gunji}, {Hayashida}, {Heyl}, {Iwakiri}, {Jorstad}, {Karas},
  {Kolodziejczak}, {La Monaca}, {Liodakis}, {Maldera}, {Manfreda}, {Marscher},
  {Marshall}, {Mitsuishi}, {Ng}, {O{\textquoteright}Dell}, {Oppedisano},
  {Papitto}, {Pavlov}, {Peirson}, {Perri}, {Pesce-Rollins}, {Pilia},
  {Possenti}, {Puccetti}, {Ramsey}, {Romani}, {Sgr{\`o}}, {Slane}, {Spandre},
  {Tamagawa}, {Tavecchio}, {Taverna}, {Tawara}, {Thomas}, {Trois}, {Tsygankov},
  {Turolla}, {Vink}, {Wu}, {Xie}, \& {Zane}}]{2022Sci...378..650K}
{Krawczynski}, H., {Muleri}, F., {Dov{\v{c}}iak}, M., {et~al.} 2022, Science,
  378, 650, \dodoi{10.1126/science.add5399}

\bibitem[{{Kuulkers} {et~al.}(1998){Kuulkers}, {Wijnands}, {Belloni},
  {M{\'e}ndez}, {van der Klis}, \& {van Paradijs}}]{1998ApJ...494..753K}
{Kuulkers}, E., {Wijnands}, R., {Belloni}, T., {et~al.} 1998, \apj, 494, 753,
  \dodoi{10.1086/305248}

\bibitem[{{Li} {et~al.}(2005){Li}, {Zimmerman}, {Narayan}, \&
  {McClintock}}]{2005ApJS..157..335L}
{Li}, L.-X., {Zimmerman}, E.~R., {Narayan}, R., \& {McClintock}, J.~E. 2005,
  \apjs, 157, 335, \dodoi{10.1086/428089}

\bibitem[{{Loktev} {et~al.}(2022){Loktev}, {Veledina}, \&
  {Poutanen}}]{Loktev22}
{Loktev}, V., {Veledina}, A., \& {Poutanen}, J. 2022, \aap, 660, A25,
  \dodoi{10.1051/0004-6361/202142360}

\bibitem[{{Loskutov} \& {Sobolev}(1979)}]{Losob79}
{Loskutov}, V.~M., \& {Sobolev}, V.~V. 1979, Astrofizika, 15, 241

\bibitem[{{Loskutov} \& {Sobolev}(1981)}]{LoskutovSobolev1981}
---. 1981, Astrofizika, 17, 97

\bibitem[{{Madsen} {et~al.}(2022){Madsen}, {Forster}, {Grefenstette},
  {Harrison}, \& {Miyasaka}}]{Madsen2022}
{Madsen}, K.~K., {Forster}, K., {Grefenstette}, B., {Harrison}, F.~A., \&
  {Miyasaka}, H. 2022, Journal of Astronomical Telescopes, Instruments, and
  Systems, 8, 034003, \dodoi{10.1117/1.JATIS.8.3.034003}

\bibitem[{{Matsuoka} {et~al.}(2009){Matsuoka}, {Kawasaki}, {Ueno}, {Tomida},
  {Kohama}, {Suzuki}, {Adachi}, {Ishikawa}, {Mihara}, {Sugizaki}, {Isobe},
  {Nakagawa}, {Tsunemi}, {Miyata}, {Kawai}, {Kataoka}, {Morii}, {Yoshida},
  {Negoro}, {Nakajima}, {Ueda}, {Chujo}, {Yamaoka}, {Yamazaki}, {Nakahira},
  {You}, {Ishiwata}, {Miyoshi}, {Eguchi}, {Hiroi}, {Katayama}, \&
  {Ebisawa}}]{maxi}
{Matsuoka}, M., {Kawasaki}, K., {Ueno}, S., {et~al.} 2009, \pasj, 61, 999,
  \dodoi{10.1093/pasj/61.5.999}

\bibitem[{{McClintock} \& {Remillard}(2006)}]{2006csxs.book..157M}
{McClintock}, J.~E., \& {Remillard}, R.~A. 2006, in Cambridge Astrophysics
  Series, Vol.~39, Compact stellar X-ray sources, ed. W.~{Lewin} \& M.~{van der
  Klis} (Cambridge: Cambridge University Press), 157--213,
  \dodoi{10.48550/arXiv.astro-ph/0306213}

\bibitem[{{Nasa High Energy Astrophysics Science Archive Research Center
  (Heasarc)}(2014)}]{2014ascl.soft08004N}
{Nasa High Energy Astrophysics Science Archive Research Center (Heasarc)}.
  2014, {HEAsoft: Unified Release of FTOOLS and XANADU}, Astrophysics Source
  Code Library, record ascl:1408.004.
\newblock \doeprint{1408.004}

\bibitem[{{Pahari} {et~al.}(2018){Pahari}, {Bhattacharyya}, {Rao},
  {Bhattacharya}, {Vadawale}, {Dewangan}, {McHardy}, {Gandhi}, {Corbel},
  {Schulz}, \& {Altamirano}}]{2018ApJ...867...86P}
{Pahari}, M., {Bhattacharyya}, S., {Rao}, A.~R., {et~al.} 2018, \apj, 867, 86,
  \dodoi{10.3847/1538-4357/aae53b}

\bibitem[{{Parmar} {et~al.}(1986){Parmar}, {Stella}, \&
  {White}}]{1986ApJ...304..664P}
{Parmar}, A.~N., {Stella}, L., \& {White}, N.~E. 1986, \apj, 304, 664,
  \dodoi{10.1086/164204}

\bibitem[{{Podgorny} {et~al.}(2023){Podgorny}, {Marra}, {Muleri}, {Rodriguez
  Cavero}, {Ratheesh}, {Dovciak}, {Mikusincova}, {Brigitte}, {Steiner},
  {Veledina}, {Bianchi}, {Krawczynski}, {Svoboda}, {Kaaret}, {Matt}, {Garcia},
  {Petrucci}, {Lutovinov}, {Semena}, {Di Marco}, {Negro}, {Weisskopf},
  {Ingram}, {Poutanen}, {Beheshtipour}, {Chun}, {Hu}, {Mizuno}, {Sixuan},
  {Tombesi}, {Zane}, {Agudo}, {Antonelli}, {Bachetti}, {Baldini},
  {Baumgartner}, {Bellazzini}, {Bongiorno}, {Bonino}, {Brez}, {Bucciantini},
  {Capitanio}, {Castellano}, {Cavazzuti}, {Chen}, {Ciprini}, {Costa}, {De
  Rosa}, {Del Monte}, {Di Gesu}, {Di Lalla}, {Donnarumma}, {Doroshenko},
  {Ehlert}, {Enoto}, {Evangelista}, {Fabiani}, {Ferrazzoli}, {Gunji},
  {Hayashida}, {Heyl}, {Iwakiri}, {Jorstad}, {Karas}, {Kislat}, {Kitaguchi},
  {Kolodziejczak}, {La Monaca}, {Latronico}, {Liodakis}, {Maldera}, {Manfreda},
  {Marin}, {Marinucci}, {Marscher}, {Marshall}, {Massaro}, {Mitsuishi}, {Ng},
  {O'Dell}, {Omodei}, {Oppedisano}, {Papitto}, {Pavlov}, {Peirson}, {Perri},
  {Pesce-Rollins}, {Pilia}, {Possenti}, {Puccetti}, {Ramsey}, {Rankin},
  {Roberts}, {Romani}, {Sgro}, {Slane}, {Soffitta}, {Spandre}, {Swartz},
  {Tamagawa}, {Tavecchio}, {Taverna}, {Tawara}, {Tennant}, {Thomas}, {Trois},
  {Tsygankov}, {Turolla}, {Vink}, {Wu}, \& {Xie}}]{2023arXiv230312034P}
{Podgorny}, J., {Marra}, L., {Muleri}, F., {et~al.} 2023, arXiv e-prints,
  arXiv:2303.12034, \dodoi{10.48550/arXiv.2303.12034}

\bibitem[{{Poutanen}(1994)}]{Poutanen1994}
{Poutanen}, J. 1994, \apjs, 92, 607, \dodoi{10.1086/192024}

\bibitem[{{Poutanen} \& {Svensson}(1996)}]{PS96}
{Poutanen}, J., \& {Svensson}, R. 1996, \apj, 470, 249, \dodoi{10.1086/177865}

\bibitem[{{Poutanen} {et~al.}(2023){Poutanen}, {Veledina}, \&
  {Beloborodov}}]{Poutanen2023}
{Poutanen}, J., {Veledina}, A., \& {Beloborodov}, A.~M. 2023, \apjl, in press,
  arXiv:2302.11674, \dodoi{10.48550/arXiv.2302.11674}

\bibitem[{{Ratheesh} {et~al.}(2023){Ratheesh}, {Dov{\v{c}}iak}, {Krawczynski},
  {Podgorn{\'y}}, {Marra}, {Veledina}, {Suleimanov}, {Rodriguez Cavero},
  {Steiner}, {Svoboda}, {Marinucci}, {Bianchi}, {Negro}, {Matt}, {Tombesi},
  {Poutanen}, {Ingram}, {Taverna}, {West}, {Karas}, {Ursini}, {Soffitta},
  {Capitanio}, {Viscolo}, {Manfreda}, {Muleri}, {Parra}, {Beheshtipour},
  {Chun}, {Cibrario}, {Di Lalla}, {Fabiani}, {Hu}, {Kaaret}, {Loktev},
  {Miku{\v{s}}incov{\'a}}, {Mizuno}, {Omodei}, {Petrucci}, {Puccetti},
  {Rankin}, {Zane}, {Zhang}, {Agudo}, {Antonelli}, {Bachetti}, {Baldini},
  {Baumgartner}, {Bellazzini}, {Bongiorno}, {Bonino}, {Brez}, {Bucciantini},
  {Castellano}, {Cavazzuti}, {Chen}, {Ciprini}, {Costa}, {De Rosa}, {Del
  Monte}, {Di Gesu}, {Di Marco}, {Donnarumma}, {Doroshenko}, {Ehlert}, {Enoto},
  {Evangelista}, {Ferrazzoli}, {Garcia}, {Gunji}, {Hayashida}, {Heyl},
  {Iwakiri}, {Jorstad}, {Kislat}, {Kitaguchi}, {Kolodziejczak}, {La Monaca},
  {Latronico}, {Liodakis}, {Maldera}, {Marin}, {Marscher}, {Marshall},
  {Massaro}, {Mitsuishi}, {Ng}, {O'Dell}, {Oppedisano}, {Papitto}, {Pavlov},
  {Peirson}, {Perri}, {Pesce-Rollins}, {Pilia}, {Possenti}, {Ramsey},
  {Roberts}, {Romani}, {Sgr{\`o}}, {Slane}, {Spandre}, {Swartz}, {Tamagawa},
  {Tavecchio}, {Tawara}, {Tennant}, {Thomas}, {Trois}, {Tsygankov}, {Turolla},
  {Vink}, {Weisskopf}, {Wu}, \& {Xie}}]{Ratheesh2023}
{Ratheesh}, A., {Dov{\v{c}}iak}, M., {Krawczynski}, H., {et~al.} 2023, Nature
  Astronomy, submitted, arXiv:2304.12752, \dodoi{10.48550/arXiv.2304.12752}

\bibitem[{{Remillard} \& {McClintock}(2006)}]{2006ARA&A..44...49R}
{Remillard}, R.~A., \& {McClintock}, J.~E. 2006, \araa, 44, 49,
  \dodoi{10.1146/annurev.astro.44.051905.092532}

\bibitem[{{Schnittman} \& {Krolik}(2009)}]{2009ApJ...701.1175S}
{Schnittman}, J.~D., \& {Krolik}, J.~H. 2009, \apj, 701, 1175,
  \dodoi{10.1088/0004-637X/701/2/1175}

\bibitem[{{Schnittman} \& {Krolik}(2010)}]{2010ApJ...712..908S}
---. 2010, \apj, 712, 908, \dodoi{10.1088/0004-637X/712/2/908}

\bibitem[{{Sobolev}(1949)}]{sob49}
{Sobolev}, V.~V. 1949, Uch. Zap. Leningrad Univ., 16

\bibitem[{{Sobolev}(1963)}]{Sob63}
---. 1963, {A treatise on radiative transfer} (Princeton: Van Nostrand)

\bibitem[{{Taverna} {et~al.}(2021){Taverna}, {Marra}, {Bianchi},
  {Dov{\v{c}}iak}, {Goosmann}, {Marin}, {Matt}, \&
  {Zhang}}]{2021MNRAS.501.3393T}
{Taverna}, R., {Marra}, L., {Bianchi}, S., {et~al.} 2021, \mnras, 501, 3393,
  \dodoi{10.1093/mnras/staa3859}

\bibitem[{{Tomsick} {et~al.}(2005){Tomsick}, {Corbel}, {Goldwurm}, \&
  {Kaaret}}]{2005ApJ...630..413T}
{Tomsick}, J.~A., {Corbel}, S., {Goldwurm}, A., \& {Kaaret}, P. 2005, \apj,
  630, 413, \dodoi{10.1086/431896}

\bibitem[{{Tomsick} {et~al.}(2014){Tomsick}, {Yamaoka}, {Corbel}, {Kalemci},
  {Migliari}, \& {Kaaret}}]{2014ApJ...791...70T}
{Tomsick}, J.~A., {Yamaoka}, K., {Corbel}, S., {et~al.} 2014, \apj, 791, 70,
  \dodoi{10.1088/0004-637X/791/1/70}

\bibitem[{{Veledina} {et~al.}(2023){Veledina}, {Muleri}, {Poutanen},
  {Podgorn{\'y}}, {Dov{\v{c}}iak}, {Capitanio}, {Churazov}, {De Rosa}, {Di
  Marco}, {Forsblom}, {Kaaret}, {Krawczynski}, {La Monaca}, {Loktev},
  {Lutovinov}, {Molkov}, {Mushtukov}, {Ratheesh}, {Rodriguez Cavero},
  {Steiner}, {Sunyaev}, {Tsygankov}, {Zdziarski}, {Bianchi}, {Bright},
  {Bursov}, {Costa}, {Egron}, {Garcia}, {Green}, {Gurwell}, {Ingram}, {Kajava},
  {Kale}, {Kraus}, {Malyshev}, {Marin}, {Matt}, {McCollough}, {Mereminskiy},
  {Nizhelsky}, {Piano}, {Pilia}, {Pittori}, {Rao}, {Righini}, {Soffitta},
  {Shevchenko}, {Svoboda}, {Tombesi}, {Trushkin}, {Tsybulev}, {Ursini},
  {Weisskopf}, {Wu}, {Agudo}, {Antonelli}, {Bachetti}, {Baldini},
  {Baumgartner}, {Bellazzini}, {Bongiorno}, {Bonino}, {Brez}, {Bucciantini},
  {Castellano}, {Cavazzuti}, {Chen}, {Ciprini}, {Del Monte}, {Di Gesu}, {Di
  Lalla}, {Donnarumma}, {Doroshenko}, {Ehlert}, {Enoto}, {Evangelista},
  {Fabiani}, {Ferrazzoli}, {Gunji}, {Hayashida}, {Heyl}, {Iwakiri}, {Jorstad},
  {Karas}, {Kislat}, {Kitaguchi}, {Kolodziejczak}, {Latronico}, {Liodakis},
  {Maldera}, {Manfreda}, {Marinucci}, {Marscher}, {Marshall}, {Massaro},
  {Mitsuishi}, {Mizuno}, {Negro}, {Ng}, {O'Dell}, {Omodei}, {Oppedisano},
  {Papitto}, {Pavlov}, {Peirson}, {Perri}, {Pesce-Rollins}, {Petrucci},
  {Possenti}, {Puccetti}, {Ramsey}, {Rankin}, {Roberts}, {Romani}, {Sgr{\`o}},
  {Slane}, {Spandre}, {Swartz}, {Tamagawa}, {Tavecchio}, {Taverna}, {Tawara},
  {Tennant}, {Thomas}, {Trois}, {Turolla}, {Vink}, {Xie}, \&
  {Zane}}]{Veledina_2023}
{Veledina}, A., {Muleri}, F., {Poutanen}, J., {et~al.} 2023, arXiv e-prints,
  arXiv:2303.01174, \dodoi{10.48550/arXiv.2303.01174}

\bibitem[{{Wang} {et~al.}(2021){Wang}, {Mastroserio}, {Kara}, {Garc{\'\i}a},
  {Ingram}, {Connors}, {van der Klis}, {Dauser}, {Steiner}, {Buisson}, {Homan},
  {Lucchini}, {Fabian}, {Bright}, {Fender}, {Cackett}, \&
  {Remillard}}]{2021ApJ...910L...3W}
{Wang}, J., {Mastroserio}, G., {Kara}, E., {et~al.} 2021, \apjl, 910, L3,
  \dodoi{10.3847/2041-8213/abec79}

\bibitem[{{Wang} {et~al.}(2022){Wang}, {Kara}, {Lucchini}, {Ingram}, {van der
  Klis}, {Mastroserio}, {Garc{\'\i}a}, {Dauser}, {Connors}, {Fabian},
  {Steiner}, {Remillard}, {Cackett}, {Uttley}, \&
  {Altamirano}}]{2022ApJ...930...18W}
{Wang}, J., {Kara}, E., {Lucchini}, M., {et~al.} 2022, \apj, 930, 18,
  \dodoi{10.3847/1538-4357/ac6262}

\bibitem[{{Weisskopf} {et~al.}(2022){Weisskopf}, {Soffitta}, {Baldini},
  {Ramsey}, {O'Dell}, {Romani}, {Matt}, {Deininger}, {Baumgartner},
  {Bellazzini}, {Costa}, {Kolodziejczak}, {Latronico}, {Marshall}, {Muleri},
  {Bongiorno}, {Tennant}, {Bucciantini}, {Dovciak}, {Marin}, {Marscher},
  {Poutanen}, {Slane}, {Turolla}, {Kalinowski}, {Di Marco}, {Fabiani},
  {Minuti}, {La Monaca}, {Pinchera}, {Rankin}, {Sgro'}, {Trois}, {Xie},
  {Alexander}, {Allen}, {Amici}, {Andersen}, {Antonelli}, {Antoniak},
  {Attina'}, {Barbanera}, {Bachetti}, {Baggett}, {Bladt}, {Brez}, {Bonino},
  {Boree}, {Borotto}, {Breeding}, {Brienza}, {Bygott}, {Caporale}, {Cardelli},
  {Carpentiero}, {Castellano}, {Castronuovo}, {Cavalli}, {Cavazzuti},
  {Ceccanti}, {Centrone}, {Citraro}, {D'Amico}, {D'Alba}, {Di Gesu}, {Del
  Monte}, {Dietz}, {Di Lalla}, {Di Persio}, {Dolan}, {Donnarumma},
  {Evangelista}, {Ferrant}, {Ferrazzoli}, {Ferrie}, {Footdale}, {Forsyth},
  {Foster}, {Garelick}, {Gunji}, {Gurnee}, {Head}, {Hibbard}, {Johnson},
  {Kelly}, {Kilaru}, {Lefevre}, {Le Roy}, {Loffredo}, {Lorenzi}, {Lucchesi},
  {Maddox}, {Magazzu}, {Maldera}, {Manfreda}, {Mangraviti}, {Marengo},
  {Marrocchesi}, {Massaro}, {Mauger}, {McCracken}, {McEachen}, {Mize}, {Mereu},
  {Mitchell}, {Mitsuishi}, {Morbidini}, {Mosti}, {Nasimi}, {Negri}, {Negro},
  {Nguyen}, {Nitschke}, {Nuti}, {Onizuka}, {Oppedisano}, {Orsini}, {Osborne},
  {Pacheco}, {Paggi}, {Painter}, {Pavelitz}, {Pentz}, {Piazzolla}, {Perri},
  {Pesce-Rollins}, {Peterson}, {Pilia}, {Profeti}, {Puccetti}, {Ranganathan},
  {Ratheesh}, {Reedy}, {Root}, {Rubini}, {Ruswick}, {Sanchez}, {Sarra},
  {Santoli}, {Scalise}, {Sciortino}, {Schroeder}, {Seek}, {Sosdian}, {Spandre},
  {Speegle}, {Tamagawa}, {Tardiola}, {Tobia}, {Thomas}, {Valerie}, {Vimercati},
  {Walden}, {Weddendorf}, {Wedmore}, {Welch}, {Zanetti}, \& {Zanetti}}]{ixpe}
{Weisskopf}, M.~C., {Soffitta}, P., {Baldini}, L., {et~al.} 2022, Journal of
  Astronomical Telescopes, Instruments, and Systems, 8, 026002,
  \dodoi{10.1117/1.JATIS.8.2.026002}

\bibitem[{{West} \& {Krawczynski}(2023)}]{West23}
{West}, A., \& {Krawczynski}, H. 2023, \apj

\bibitem[{{Wilms} {et~al.}(2000){Wilms}, {Allen}, \&
  {McCray}}]{2000ApJ...542..914W}
{Wilms}, J., {Allen}, A., \& {McCray}, R. 2000, \apj, 542, 914,
  \dodoi{10.1086/317016}

\bibitem[{{Yan} \& {Wang}(2011)}]{2011RAA....11..631Y}
{Yan}, L.-H., \& {Wang}, J.-C. 2011, Research in Astronomy and Astrophysics,
  11, 631, \dodoi{10.1088/1674-4527/11/6/002}

\bibitem[{{Zdziarski} \& {Gierli{\'n}ski}(2004)}]{Zdziarski2004}
{Zdziarski}, A.~A., \& {Gierli{\'n}ski}, M. 2004, Progress of Theoretical
  Physics Supplement, 155, 99, \dodoi{10.1143/PTPS.155.99}

\bibitem[{{Zdziarski} {et~al.}(2001){Zdziarski}, {Grove}, {Poutanen}, {Rao}, \&
  {Vadawale}}]{Zdziarski2001}
{Zdziarski}, A.~A., {Grove}, J.~E., {Poutanen}, J., {Rao}, A.~R., \&
  {Vadawale}, S.~V. 2001, \apjl, 554, L45, \dodoi{10.1086/320932}

\bibitem[{{Zdziarski} {et~al.}(1996){Zdziarski}, {Johnson}, \&
  {Magdziarz}}]{1996MNRAS.283..193Z}
{Zdziarski}, A.~A., {Johnson}, W.~N., \& {Magdziarz}, P. 1996, \mnras, 283,
  193, \dodoi{10.1093/mnras/283.1.193}

\bibitem[{{{\.Z}ycki} {et~al.}(1999){{\.Z}ycki}, {Done}, \&
  {Smith}}]{1999MNRAS.309..561Z}
{{\.Z}ycki}, P.~T., {Done}, C., \& {Smith}, D.~A. 1999, \mnras, 309, 561,
  \dodoi{10.1046/j.1365-8711.1999.02885.x}

\bibitem[{{{\.Z}ycki} {et~al.}(2001){{\.Z}ycki}, {Done}, \&
  {Smith}}]{Zycki2001}
---. 2001, \mnras, 326, 1367, \dodoi{10.1111/j.1365-2966.2001.04698.x}

\end{thebibliography}
\bibliographystyle{aasjournal}

\appendix
\section{Data Reduction}
\label{sec:datareduction}

{\it IXPE} \citep{ixpe} observed 4U\,1630--47 for $\sim141$ ksec between 2023 March 10 19:21:04 UTC and 2023~March 13 19:02:48 UTC under observation ID 02250601. The {\it IXPE} processed level-2 data was obtained from the HEASARC archive. The data are publically available from the web-site \url{https://heasarc.gsfc.nasa.gov/docs/ixpe/archive/}. The analysis of the {\it IXPE} data was performed using the {\sc ixpeobssim} software version 28.4.0 \citep{Baldini2022} based on the level-2 processed data. The source region was chosen in SAOImage DS9 software \citep{ds9} as a circular area with a 60\arcsec\  radius centered at ($16^\text{h}34^\text{m}03^\text{s}.3$, $-47\degr 23$\arcmin 16\arcsec.8). We did not extract the background due to possible contamination of source photons \citep{2023AJ....165..143D}. 
The PD and PA were computed using the {\tt PCUBE} algorithm 
incorporated in {\sc ixpeobssim} to calculate the polarization signature of the observation independent of a model. 
Version 11 of the {\it IXPE} response functions were used to process the data.

NICER \citep{nicer} is a soft X-ray spectral-timing instrument covering the 0.2--12 keV energy band. NICER observed the source from 2023 March 10 18:09:24 UTC to March 13 19:35:58 UTC under observation IDs 6557010XXX (101,102,201,202,301,302) for a total of $\sim 32.81$~ks of useful time among the 6 observations. The data were reduced using {\sc nicerdas} v10 software and the {xti20221001} release of NICER CALDB products.  The SCORPEON background model was adopted.  Observations were filtered for hot detectors, corrected for detector deadtime, and screened to remove candidate good-time intervals with substantially elevated background or candidate intervals less than 100~s long.  In the paper we show the NICER spectra combined according to Period~1 or Period~2 of the {\it IXPE} observation. We have used {\tt addspec} tool from the {\sc heasoft} package \citep{2014ascl.soft08004N} for this purpose.

The {\it NuSTAR} \citep{nustar} spacecraft observed the source under observation IDs $8090231300\#$(2,4,6) and collected a total of $\sim 28.35$~ks of net exposure. The data were processed with the {\tt NuSTARDAS} software (version 2.1.2) of the {\sc heasoft} package (version 6.31.1). The source events were selected with a circular region of 60\arcsec\ radii centered at the source coordinates ($16^\text{h}34^\text{m}01^\text{s}.6101$, $-47\degr 23$\arcmin 34\arcsec.806) for both focal plane modules (FPMA/FPMB). The background region was selected as circular region with radius $\sim$91.6\arcsec\ centered at ($16^\text{h}34^\text{m}46^\text{s}.6422$, $-47\degr 24$\arcmin 03\arcsec.752). The first observation was taken during the Period 1 with lower flux, while the other two observations correspond to the Period 2. We compared the spectra and combined the 2nd and 3rd spectra for each FPM using \texttt{addspec} from {\sc ftool}  to have one representative spectra per FPM for the Period 2.

\section{Spectral Fit}
\label{sec:spectralfit}

\begin{figure}[ht!]
\epsscale{1.1}
\hspace*{-2cm}a)\hspace{0.4\textwidth}b)\\
\plotone{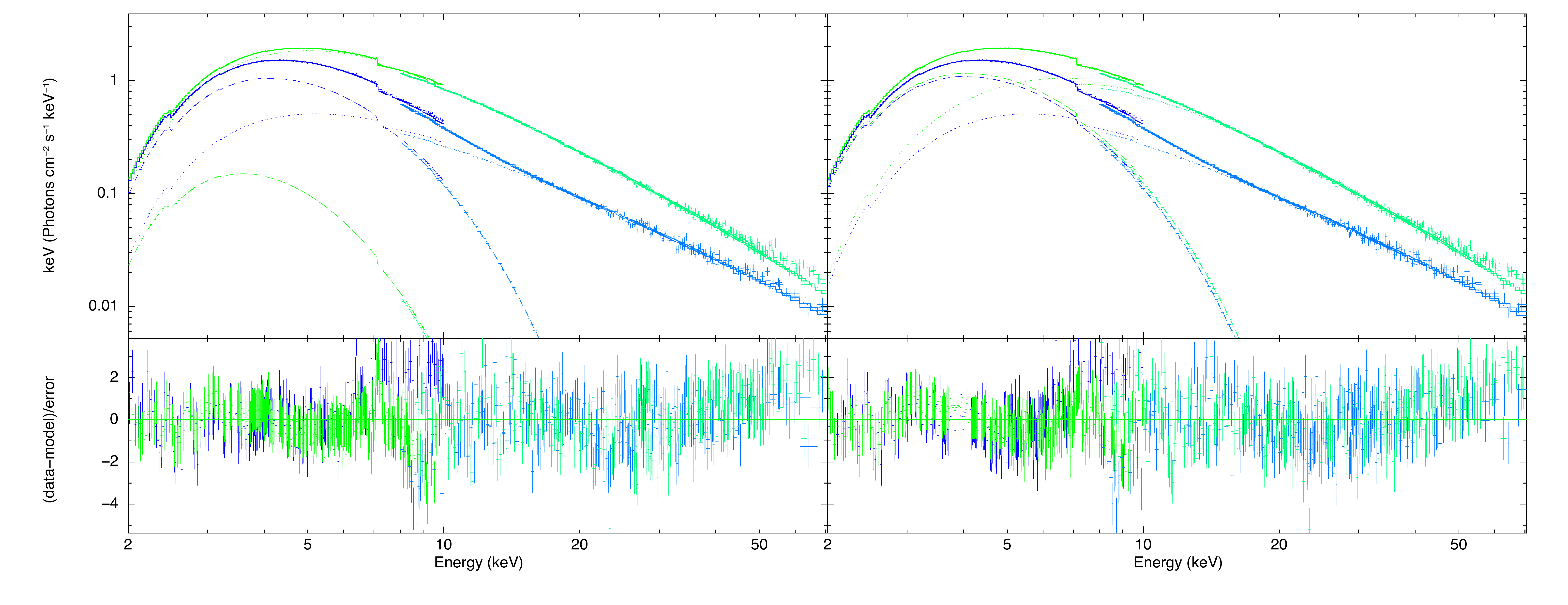}
\caption{
Fits of 4U\,1630--47 NICER and {\it NuSTAR} X-ray spectra for Period 1 (blue) and Period 2 (green): a) Disk blackbody assumed as seed radiation for the power-law component (Fit 1). b) Single temperature blackbody assumed as seed radiation for the power-law component (Fit 2). Unfolded spectra around the best-fitting model in $F_E$ representation, the total model (solid) and the {\tt kerrbb} (dashed) and {\tt nthcomp} (dotted) contributions for each data set are shown in the top panels while the data-model residuals in $\sigma$ are shown in the bottom panels.
\label{fig:specfits}}
\end{figure}

In order to study the polarization properties of the thermal and Comptonized component we performed a joint fit on the NICER and {\it NuSTAR} spectra of the SPL state observation. Since our aim here is only to give an estimate of the polarization degree and angle of the two spectral components, we performed our analysis on the two instruments' time-averaged spectra, subdivided into two groups each, corresponding to Periods 1 and 2 of the SPL state {\it IXPE} observation. 
Moreover, to further simplify our approach, we restricted our study to the 2--10 keV energy range for NICER data and to the 8--70 keV range for {\it NuSTAR} ones in order to reduce cross-calibration uncertainties between the two instruments. The choice to analyze {\it NuSTAR} data starting from 8 keV, in particular, belongs to the large inconsistencies between the NICER and {\it NuSTAR} data below this energy---although some cross-calibration residuals can still be observed in the 8--9~keV range. We used the {\it NuSTAR} spectra up to 70~keV since above that the background was comparable to the data. A complete, time-resolved spectral analysis, as well as the study of the low energy inconsistencies between NICER and {\it NuSTAR} data, is beyond the scope of this paper and will be addressed in a future publication.
We used the \textsc{xspec} package (v12.13.0c) and employed the following model in the analysis:
\begin{equation}
\tt edge*edge*tbabs(kerrbb+nthcomp) .
    \label{eq:SpectralFit}
\end{equation}
The model comprises of thermal thin accretion disk emission accounting for relativistic effects \citep[{\tt kerrbb},][]{2005ApJS..157..335L}, thermally Comptonized continuum emission \citep[{\tt nthcomp},][]{1996MNRAS.283..193Z,1999MNRAS.309..561Z}, and absorption by the interstellar medium \citep[{\tt tbabs},][]{2000ApJ...542..914W}. Following the approach from Paper\,I, we fixed the distance of the source in the {\tt kerrbb} model to the value of $D=11.5$ kpc; moreover, we kept the system inclination fixed at the value of $i=75\degr$, leaving only the black hole spin, mass, and accretion rate free to vary in the fitting procedure. 
A {\tt cloudy} \citep{2017RMxAA..53..385F} absorption table was used in Paper\,I to model the absorption lines detected in the observation of the source in HSS, likely produced by a highly-ionized outflowing plasma (i.e. with ionization parameter $\xi \approx 10^5$ and hydrogen column density $N_{H} \approx 10^{24}\ \mathrm{cm}^{-2}$).
If we use the {\tt cloudy} component and assume the same ionization parameter of the HSS observation,  it is possible to obtain an upper limit of $N_{\rm H} \leq 10^{22}\ \mathrm{cm}^{-2}$ on the wind column density along the line of sight. However, if the ionization parameter is allowed to vary freely it is usually fitted to unrealistically high values. Additionally, the SPL state observation shows no prominent absorption lines so this component was no longer used in the fitting procedure.
We used the {\tt nthcomp} component assuming either disk blackbody or blackbody seed radiation. For Fit~1, we assumed multicolor disk blackbody seed radiation (inp\_type parameter = 1) and fixed its temperature to the values obtained from initial modeling using {\tt diskbb} ($kT_{\rm bb}=1.46 \substack{+0.02 \\ -0.01};\ 1.54 \substack{+0.01 \\ -0.02}$ keV in Period 1 and 2, respectively). For Fit~2, we used a single blackbody as the input radiation (inp\_type parameter = 0) and instead left the temperature free to vary in the fitting procedure.
The {\tt nthcomp} input radiation modified the fluxes contributions, as presented in Table \ref{tab:PD-PA}, and consequently the polarization properties of the power-law component. This is due to the different low energy contributions of {\tt nthcomp} when using a multicolor black body in place of a single black body, which influences the {\tt kerrbb} accretion rate in the fitting procedure and consequently the thermal radiation contribution to the total flux.
Figure \ref{fig:specfits} shows the unfolded spectra and data residuals for both fits. The Period 2 {\tt kerrbb} contribution to the total flux in Fit 2 is significantly larger than in Fit 1 as denoted by the dashed green lines.

Additionally, following Paper\,I, an empirical absorption {\tt edge} model was used at $2.42$ and $9.51$ keV to account for reported instrumental features in the NICER and {\it NuSTAR} spectra, respectively \citep{2021ApJ...910L...3W,2023arXiv230312034P}. 
The cross-calibration model {\tt MBPO} employed in \citet{2022Sci...378..650K} was used to account for cross-calibration uncertainties between NICER and {\it NuSTAR} allowing the spectral slope and normalization to vary.
For the {\it NuSTAR} focal plane module A (FPMA) we fixed the normalization to 1 for all fitting groups, corresponding to the recommended value in \citet{Madsen2022} and kept the slope fixed to zero.
For the fit presented in Table \ref{t:spectralfitparams} we obtained the normalization values of $1.035 \pm 0.002$ and $0.994 \pm 0.001$, and the slope values of $0.0664 \pm 0.0033$ and $0.0095 \pm 0.0025$, for the NICER and {\it NuSTAR} FPMB observations, respectively. The best-fit parameters of this analysis are shown in Table \ref{t:spectralfitparams} for a $\chi^2/\mathrm{dof} = 2502.68/2399$, when using a disk blackbody input radiation for the {\tt nthcomp} component, and a $\chi^2/\mathrm{dof} = 2470.75/2399$ assuming a blackbody input for the power-law component. It is worth noting that in our simplified approach the data are consistently above the model in the high energy tail of the spectra (45--70~keV) with both models further motivating the need for a more detailed analysis of the spectral properties of this source.

As a following step, we included {\it IXPE} spectra in the fitting procedure, dividing them into two groups corresponding to the periods of the SPL state observation. We allowed all the parameters of the {\tt MBPO} model to vary independently for each of the three {\it IXPE} detector units in both periods.
For the fits presented in Table \ref{t:spectralfitparams}, we found the $\Delta \Gamma_1$ values of $0.118 \pm 0.033$, $0.073 \pm 0.032$, $0.099 \pm 0.019$, the $\Delta \Gamma_2$ values of $-0.27 \pm 0.11$, $-0.41 \pm 0.19$, $-0.77 \pm 0.31$, the $E_{\rm br}$ values of $4.43 \pm 0.38$, $4.79 \pm 0.45$, $5.46 \pm 0.34$ and the normalization values of $0.7361 \pm 0.0046$, $0.7036 \pm 0.0054$, $0.6786 \pm 0.0049$ for the {\it IXPE} DUs 1,2,3, respectively, in Period 1 of the SPL state. In Period 2, we obtained the following values for the three {\it IXPE} DUs: $\Delta \Gamma_1 = 0.013 \pm 0.011, \ 0.075 \pm 0.018, \ 0.076 \pm 0.016$; $\Delta \Gamma_2 = 0.59 \pm 0.15, \ -0.416 \pm 0.063, \ 0.458 \pm 0.097$; $E_{\rm br}=5.47 \pm 0.25, \ 4.64 \pm 0.20, 5.00 \pm 0.26$; normalization of $0.7204 \pm 0.0042, \ 0.7184 \pm 0.0033, \ 0.6843 \pm 0.0029$. We found that the best-fitting model has a $\chi^2/\mathrm{dof} = 3693.90/3293$. The reduced $\chi^2$ is greater than one even accounting for $1\%$ systematic uncertainties for the NICER data sets, within the mission's recommendations\footnote{NICER calibration recommendations can be fount at \url{https://heasarc.gsfc.nasa.gov/docs/nicer/analysis_threads/cal-recommend/}.}
This result is likely due to our simplified approach of performing a time-averaged analysis on a highly variable source.




\begin{deluxetable}{ccccccc}
\tabletypesize{\scriptsize}
\tablewidth{0pt} 
\tablecaption{Spectral fit parameters \label{t:spectralfitparams}}

\tablehead{
\colhead{Component} & \colhead{Parameter (unit)}& \colhead{Description} & \multicolumn{2}{c}{Value Fit 1} & \multicolumn{2}{c}{Value Fit 2} 
\\ & & & SPL Period 1 & SPL Period 2 & SPL Period 1 & SPL Period 2
} 
\startdata 
{\tt tbabs}& $N_\textrm{H}$ ($10^{22}$\,cm$^{-2}$) & Hydrogen column density & $7.84\substack{+,0.02 \\ -0.04}$ & $7.71\substack{+0.02 \\ -0.02}$ & $7.63\substack{+0.03 \\ -0.03}$ & $7.78\substack{+0.02 \\ -0.02}$\\
\hline
{\tt kerrbb}& $\eta$ & Disk power ratio & $0$ (frozen) & - & - & - \\
& $a$ & Black hole spin & $0.71\substack{+0.25 \\ -0.15}$ & - & $0.72\substack{+0.18 \\ -0.21}$ & - \\
& $i$ (deg) & Inclination & $75.00$ (frozen) & - & - & - \\
& $M_{\textrm{bh}}$ ($M_\odot$) & Black hole mass  & $10.51\substack{+3.51 \\ -2.54}$ & - & $9.37\substack{+2.95 \\ -2.14}$ & - \\
& $M_{\textrm{dd}}$ ($10^{18}$\,g s$^{-1}$) & Effective mass accretion rate  & $6.22\substack{+0.98 \\ -0.33}$ & $1.22\substack{+0.25 \\ -0.21}$ & $4.91\substack{+0.61 \\ -0.48}$ & $6.91\substack{+0.57 \\ -0.61e}$ \\
& $D$ (kpc) & Distance & $11.5$ (frozen) & - & - & - \\
& $hd$ & Hardening factor & $1.7$ (frozen) & - & - & - \\
& $r_\textrm{flag}$ & Self-irradiation & $1$ (frozen) & - & - & - \\
& $l_\textrm{flag}$ & Limb-darkening  & $0$ (frozen) & - & - & - \\
& norm & Normalization  & $1.0$ (frozen) & - & - & - \\
\hline
{\tt nthcomp} & $\Gamma$ & Photon index & $2.64\substack{+0.02 \\ -0.01}$ & $2.94\substack{+0.01 \\ -0.01}$ & $2.61\substack{+0.02 \\ -0.02}$ & $2.93\substack{+0.01 \\ -0.01}$ \\
& $kT_\textrm{e}$ (keV) & Electron temperature & $500.00$ (frozen) & - & - & - \\
& $kT_\textrm{bb}$ (keV) & Seed photon temperature & $1.46\substack{+0.02 \\ -0.01}$ & $1.54\substack{+0.01 \\ -0.02}$ & $0.91\substack{+0.24 \\ -0.18}$ & $1.88\substack{+0.38 \\ -0.36}$ \\
& inp\_type & Seed photon shape & $1.0$ (frozen) & - & $0.0$ (frozen) & - \\
& $z$ & Redshift & $0.0$ (frozen) & - & - & - \\
& norm & Normalization & $1.09\substack{+0.02\\-0.02}$ & $3.68\substack{+0.01 \\ -0.01}$ & $0.41\substack{+0.05 \\ -0.05}$ & $0.13\substack{+0.02 \\ -0.02}$ \\
\hline
{\tt edge} 1& edgeE (keV)& Threshold energy & $2.43\substack{+0.01 \\ -0.01}$ & - & - & - \\
& MaxTau ($10^{-2}$)& Absorption Depth at threshold energy & $6.14\substack{+0.40 \\ -0.41}$ & - & - & - \\
\hline
{\tt edge} 2& edgeE (keV) & Threshold energy & $9.49\substack{+0.05 \\ -0.05}$ & - & - & -\\
& MaxTau ($10^{-2}$)& Absorption Depth at threshold energy & $1.88\substack{+0.21 \\ -0.22}$ & - & - & - \\
\enddata
\tablecomments{Best-fitting parameters for joint NICER and {\it NuSTAR} spectral fitting for periods corresponding to Period 1 and Period 2. Uncertainties are stated at the 90\% confidence level. Parameters are calculated assuming a disk blackbody seed radiation (Fit 1) or a blackbody seed radiation (Fit 2) for the power-law component of the spectra.}.
\end{deluxetable}




\end{document}